\documentclass[sigconf]{acmart}
\renewcommand\footnotetextcopyrightpermission[1]{} %
\usepackage{booktabs}
\usepackage{url}
\usepackage{flushend}
\usepackage{graphicx}
\usepackage[font=small,labelfont=bf]{caption}

\usepackage{subcaption}
\usepackage{enumitem}
\usepackage{amssymb,amsmath}
\usepackage[T1]{fontenc}
\usepackage[hyphenbreaks]{breakurl}
\newcommand{\descr}[1]{\smallskip\noindent\textbf{#1}}
\newcommand{\argmax}{\arg\!\max}
\newcommand{\argmin}{\arg\!\min}

\makeatletter
\def\@copyrightspace{\relax}
\makeatother
\begin{document} 
\title{Hate is not Binary:\\ Studying Abusive Behavior of \#GamerGate on Twitter}

\author{Despoina Chatzakou$^{\dagger}$, Nicolas Kourtellis$^{\ddagger}$, Jeremy Blackburn$^{\ddagger}$\\[-2ex]
Emiliano De Cristofaro$^{\sharp}$, Gianluca Stringhini$^{\sharp}$, Athena Vakali$^{\dagger}$}
\affiliation{$^\dagger$Aristotle University of Thessaloniki \hspace{0.2cm} $^\ddagger$Telefonica Research \hspace{0.2cm} $^{\sharp}$University College London\\
deppych@csd.auth.gr, nicolas.kourtellis@telefonica.com, jeremy.blackburn@telefonica.com\\
e.decristofaro@ucl.ac.uk, g.stringhini@ucl.ac.uk, avakali@csd.auth.gr}

\renewcommand{\shortauthors}{D. Chatzakou et al.}

\newcommand{\nknote}[1]{\textcolor{blue}{\small \bf [NK: #1]}}

\newenvironment {squishlist}
{\begin{list}{$\bullet$}
  { \setlength{\itemsep}{1pt}
     \setlength{\parsep}{1pt}
     \setlength{\topsep}{1pt}
     \setlength{\partopsep}{1pt}
     \setlength{\leftmargin}{1.5em}
     \setlength{\labelwidth}{1em}
     \setlength{\labelsep}{0.5em} } }
{\end{list}}

\begin{abstract}
Over the past few years, online bullying and aggression have become increasingly prominent, and manifested in many different forms on social media. However, there is little work analyzing the characteristics of abusive users and what distinguishes them from typical social media users. In this paper, we start addressing this gap by analyzing tweets containing a great large amount of abusiveness. We focus on a Twitter dataset revolving around the Gamergate controversy, which led to many incidents of cyberbullying and cyberaggression on various gaming and social media platforms. We study the properties of the users tweeting about Gamergate, the content they post, and the differences in their behavior compared to typical Twitter users.

We find that while their tweets are often seemingly about aggressive and hateful subjects, ``Gamergaters'' do not exhibit common expressions of online anger, and in fact primarily differ from typical users in that their tweets are less joyful. They are also more engaged than typical Twitter users, which is an indication as to how and why this controversy is still ongoing. Surprisingly, we find that Gamergaters are less likely to be suspended by Twitter, thus we analyze their properties to identify differences from typical users and what may have led to their suspension. We perform an unsupervised machine learning analysis to detect clusters of users who, though currently active, could be considered for suspension since they exhibit similar behaviors with suspended users. Finally, we confirm the usefulness of our analyzed features by emulating the Twitter suspension mechanism with a supervised learning method, achieving very good precision and recall.
\end{abstract}

\maketitle

\section{Introduction}\label{sec:intro}

Abuse on social media is becoming a pressing issue.
Over the past few years, social networks have not only been targeted by bots and fraudsters~\cite{Wang2010SpamDetectionTwitter,benevenuto2010detecting,stringhini2010detecting}, but have also been used as a platform for harassing and trolling other individuals~\cite{sanchez2011twitter}.
Detecting and mitigating such activities presents important challenges since abuse performed by human-controlled accounts tends to be less homogeneous than the one performed by bots, making it hard to identify the characteristics that distinguish them from non-abusive attacks (and detect them).
Recent work showed that human-controlled accounts involved in harassment actually present degrees of synchronized activity~\cite{hine2016longitudinal}.
However, no systematic measurement has been performed to understand what distinguishes a social network account behaving in an abusive way from a typical one.
Such an understanding is crucial to enable effective mitigation and help social network operators to detect and block these accounts.

\descr{Roadmap.} In this paper, we start addressing this gap by performing a large-scale comparative study of abusive accounts on Twitter, aiming to understand their characteristics and how they differ from typical accounts.
We collect a large dataset of tweets related to the Gamergate (GG) controversy~\cite{chatzakou2017Gamergate}, which after two years since its start has evolved into a fairly mature, pseudo-political movement that is thought to encompass semi-organized campaigns of hate and harassment by its adherents, known as Gamergaters (GGers), against women in particular.
Then, we explore the differences between the GG-related accounts identified as abusive, and random Twitter accounts, investigate how these differences lead to disproportional suspension rates by Twitter, and discuss possible causes of these differences.
We also look at accounts of users that were deleted by their owner and not by Twitter.
Further, we cluster GG accounts that exhibit similar behavior, aiming to identify groups of similar accounts that should have been suspended by Twitter but are instead still active.
Based on the findings of our clustering, we reason about what may have driven Twitter to not suspend them.
Finally, we test the performance of a supervised method to automatically suspend Twitter users based on the various features analyzed.

\descr{Findings.} In summary, we discover that users involved in Gamergate were already-existing Twitter users probably drawn to the controversy, which might be the reason why GG exploded on Twitter in the first place.
While the subject of their tweets is seemingly aggressive and hateful, GGers do not exhibit common expressions of online anger, and in fact primarily differ from random users in that their tweets are less joyful. 
We find that despite their clearly anti-social behavior, GGers tend to have more friends and followers than random users and being more engaging in the platform may have allowed this controversy to continue until now.
Surprisingly, we find that GGers are disproportionally \emph{not suspended} from Twitter in comparison to random users, which is rather unexpected given their hateful and aggressive postings.
Suspended GG users expressed more aggressive and repulsive emotions, offensive language, and interestingly, more joy than suspended random users, and their high posting and engaging activity may have delayed their suspension from Twitter.
Also GGers who deleted their account demonstrated the most activity in comparison to other users (deleted or suspended), exhibited signs of distress, fear, and sadness.
They have probably showed these emotions through their high posting activity filled with anger, reduced joy, and negative sentiment.
Such users have small social ego-networks which may have been unsupportive or too small to help them before deleting their accounts.

\descr{Paper Organization.} The rest of the paper is organized as follows.
The next section reviews related work on measuring abusive behaviors on online platforms.
Section~\ref{sec:dataset} introduces our dataset and the steps taken for cleaning and preparing it for analysis, then,
in Section~\ref{sec:comparison}, we analyze the behavioral patterns exhibited by GGers, and compare them to random Twitter users.
In Section~\ref{sec:statuses}, we discuss how users get suspended on Twitter, differences observed between GGers and random users, reasons for deviating from the expected rates, and a basic effort to emulate Twitter's suspension mechanism.
In Section~\ref{sec:discussion} we discuss our findings and conclude.

\section{Related work}\label{sec:related-work}

We now review related work on studying/detecting offensive, abusive, aggressive or bullying content on social media sources.
Chen et al.~\cite{Chen2012DetectingOffensiveLanguage} aim to detect offensive content, as well as, potential offensive users based on YouTube comments.
Both Yahoo Finance~\cite{Djuric2015HateSpeechDetection,Nobata2016AbusiveLanguageDetection} and Yahoo Answers~\cite{kayes2015ya-abuse} have been used as a source of information for detecting hate and/or abusive content.
More specifically,~\cite{kayes2015ya-abuse} studied a Community-based Question-Answering (CQA) site and finds that users tend to flag abusive content in an overwhelmingly correct way.

Cyberbullying has also attracted a lot of attention lately, for instance~\cite{chatzakou2017websci},~\cite{Hosseinmardi2014TowardsUnderstandingCyberbullying} and~\cite{Hosseinmardi2015} focus on Twitter, Ask.fm, and Instagram, respectively, to detect existing bullying cases out of text sources.
\cite{chatzakou2017websci} considers a variety of features, i.e., user, text, and network-based, to distinguish bullies and aggressors from typical Twitter users.
In addition to text sources,~\cite{Hosseinmardi2015} also tries to associate an image's topic (e.g., drugs, celebrity, sports, etc) with cyberbullying events.
In~\cite{Dinakar2011ModelingDetectionTextualCyberbullying}, the cyberbullying phenomenon is further decomposed to specific sensitive topics, i.e., race, culture, sexuality, and intelligence, by analyzing YouTube comments extracted from controversial videos.
A study of specific cyberbullying cases, e.g., threats and insults, is also conducted in~\cite{Hee2015AutomaticDetectionPreventionCyberbullying} by considering Dutch posts extracted from Ask.fm.
Apart from cyberbullying, they also study specific user behaviors: harasser, victim, and bystander-defender or bystander-assistant who support the victim or the harasser, respectively.
In follow-up work~\cite{sanchez2011twitter}, the authors exploit Twitter messages to detect bullying cases which are specifically related to the gender bullying phenomenon.
Finally, in~\cite{Dadvar2014ExpertsMachinesAgainstBullies}, YouTube users are characterized based on a ``bulliness'' score.
The rise of cyberbullying, and abusive incidents in general, is also evident in online game communities.
Since these communities are widely used by people of all ages, such a phenomenon has attracted the interest of the research community.
For instance,~\cite{kwak2015exploringcyberbullying} studies cyberbullying and toxic behaviors in team competition online games in an effort to detect, prevent, and counter-act toxic behavior.
\cite{fox2014Sexism} investigates the prevalence of sexism in online game communities finding personality traits, demographic variables, and levels of game-play predicted sexist attitudes towards women who play video games.
Overall, previous work considers various attributes to distinguish between normal and abusive behavior, like text-based attributes, e.g., URLs and  Bag of Words (BoW), lexicon-based (offensive word dictionary), or user/activity based attributes, e.g., number of friends/followers and users' account age.
Our work aims to use such attributes to study and understand the different behavioral patterns between random and Gamergate Twitter users, while shedding light on how such differences affect their suspension and deletion rates on Twitter.

\descr{Analysis of Gamergate.}
To create an abuse-related dataset, i.e., a dataset containing abusive behavior with high probability, previous works rely on a number of words (i.e., seed words) which are highly related with the manifestation of abusive/aggressive events.
In this sense, a popular term that can serve as a seed word is the \#GamerGate hashtag which is one of the most well documented large-scale instances of bullying/aggressive behavior we are aware of~\cite{Massanari09102015}.
The Gamergate controversy stemmed from alleged improprieties in video game journalism, which quickly grew into a larger campaign centered around sexism and social justice.
With individuals on both sides of the controversy using it, and extreme cases of bullying and aggressive behavior associated with it (e.g., direct threats of rape and murder), \#GamerGate can serve as a relatively unambiguous hashtag associated with texts that are likely to involve abusive/aggressive behavior from a fairly mature, hateful online community.
In~\cite{Mortensen2016}, the author shows that \#GamerGate can be likened to hooliganism, i.e., a leisure-centered aggression were fans are organized in groups to attack another group's members.
Also,~\cite{guberman2017} aims to detect toxicity on Twitter, considering \#GamerGate to collect a sufficient number of harassment-related posts.
In this paper, we also study a number of abusive users involved in this controversy via \#GamerGate.
However, we are the first to investigate the attributes characterizing these users with respect to their Twitter account status (active, suspended, deleted), and to perform an unsupervised and supervised analysis of suspicious users for possible suspension.

\section{Dataset}\label{sec:dataset}
In this section, we present the data used throughout the rest of the paper, as well as the two prepocessing steps: spam removal and dataset cleaning.

\begin{figure*}[!t]
	\centering
	\begin{subfigure}[b]{0.24\textwidth}
		\includegraphics[width=\textwidth]{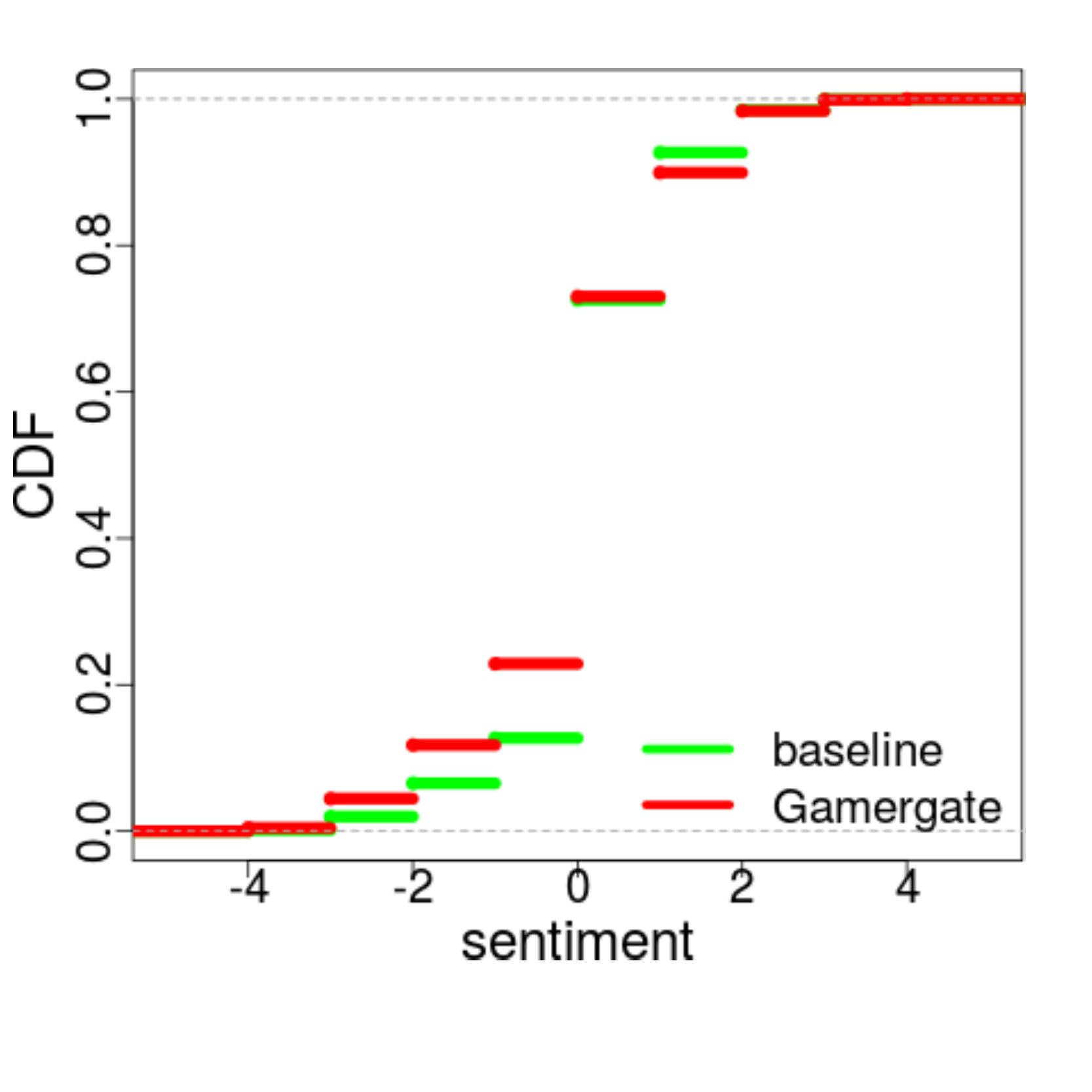}
		\captionsetup{font=scriptsize}
		\vspace{-0.85cm}
		\caption{Sentiment distribution.}
		\label{fig:baseline_hatebase_sentiment}
	\end{subfigure}
	\begin{subfigure}[b]{0.24\textwidth}
		\includegraphics[width=\textwidth]{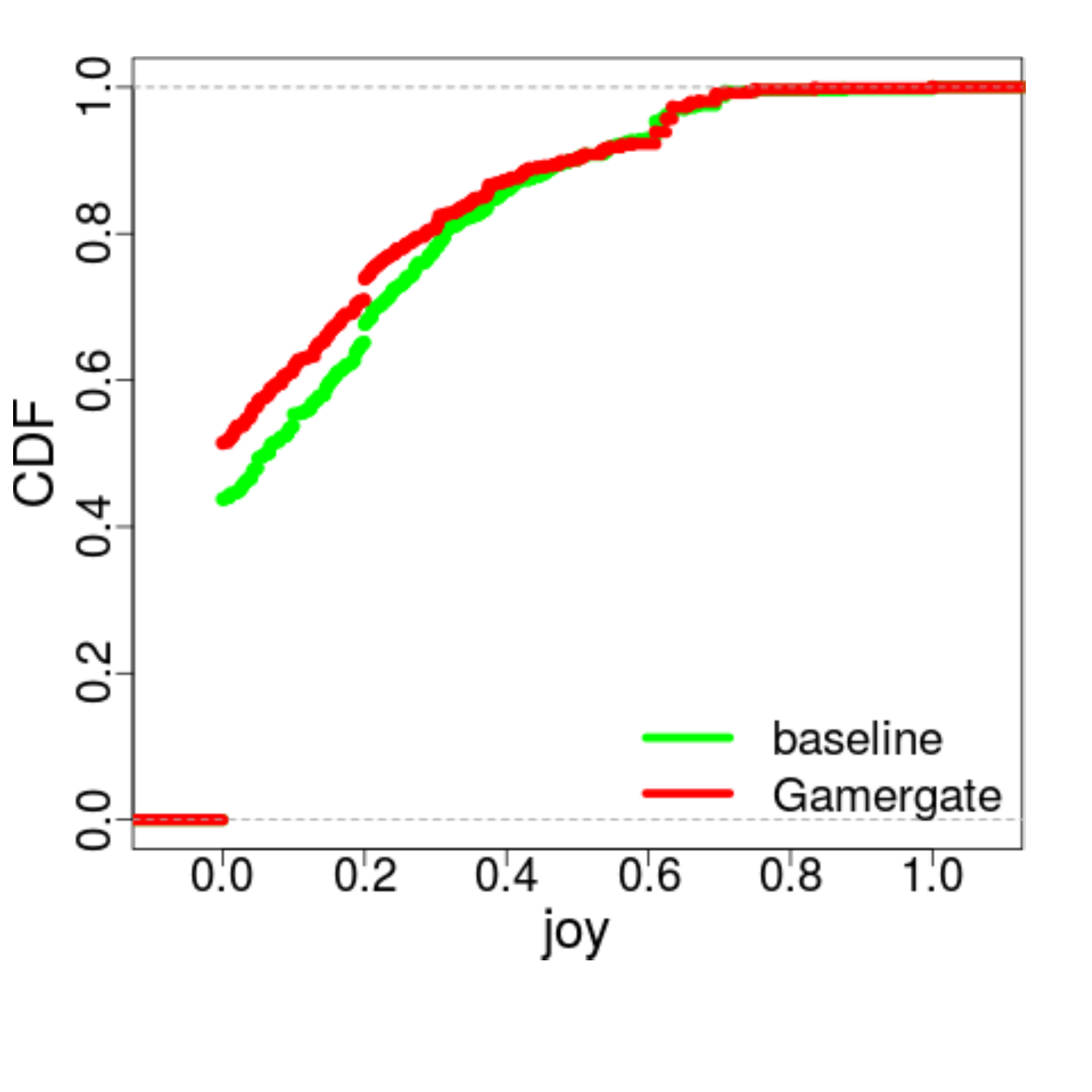}
		\captionsetup{font=scriptsize}
		\vspace{-0.85cm}
		\caption{Joy distribution.}
		\label{fig:baseline_hatebase_joy}
	\end{subfigure}	
	\begin{subfigure}[b]{0.24\textwidth}
		\includegraphics[width=\textwidth]{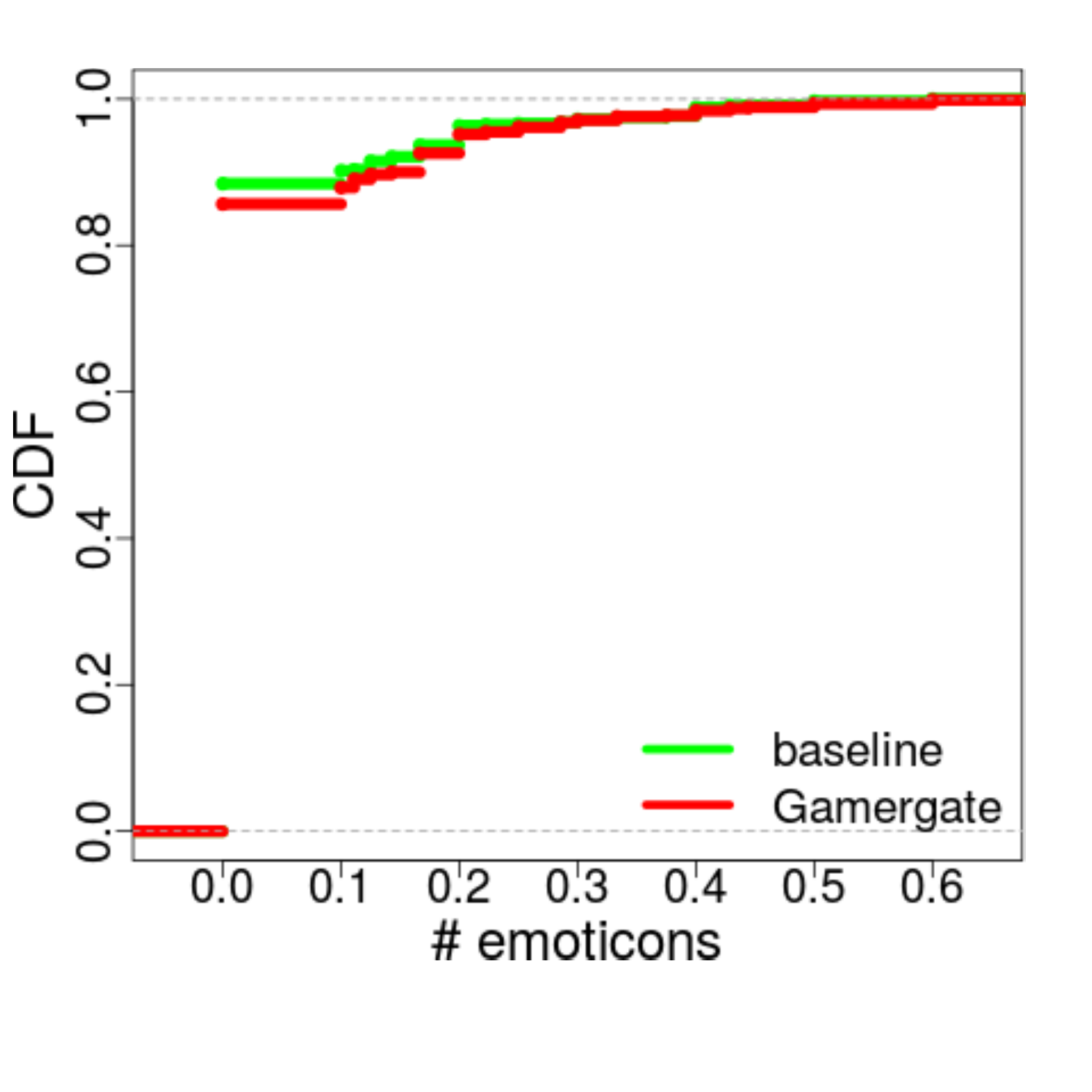}
		\captionsetup{font=scriptsize}
		\vspace{-0.85cm}
		\caption{Emoticons distribution.}
		\label{fig:baseline_hatebase_emoticons}
	\end{subfigure}
	\begin{subfigure}[b]{0.24\textwidth}
		\includegraphics[width=\textwidth]{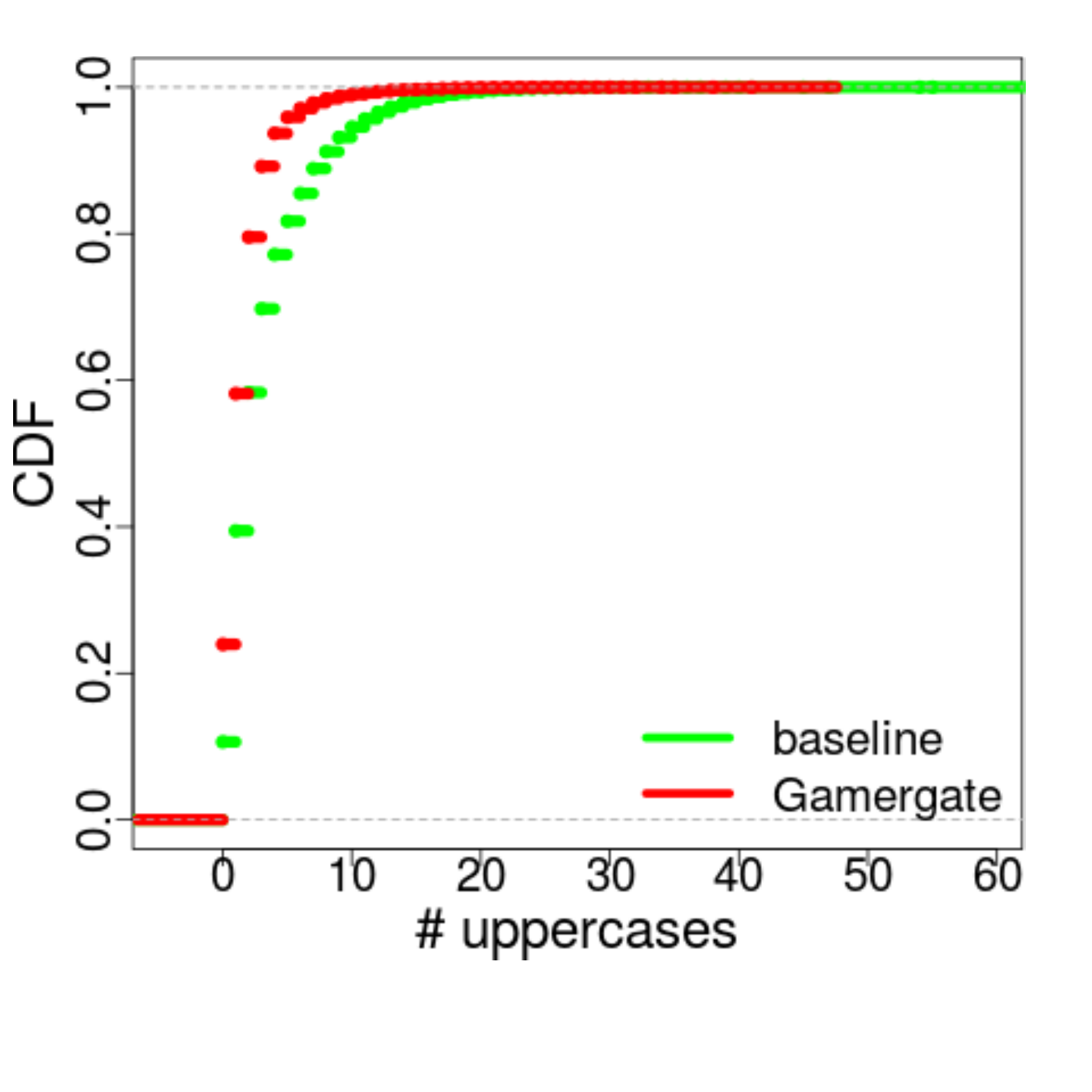}
		\captionsetup{font=scriptsize}
		\vspace{-0.85cm}
		\caption{Uppercases distribution.}
		\label{fig:baseline_hatebase_uppercases}
	\end{subfigure}	
	\vspace{-0.15cm}
	\caption{Average CDF distribution of (a) Sentiment, (b) Joy, (c) Emoticons, (d) Uppercases in baseline and Gamergate datasets.}
	\vspace{-0.2cm}
\end{figure*}	

\subsection{Data Collection}

The data used in the next sections were collected between June and August 2016 using the Twitter Streaming API~\cite{Streaming} which gives access to $1\%$ of all tweets.
Data returned from the Twitter API include either user-related info, e.g., users' follower/friends count, total number of posted, liked and favorited tweets, or text-related, e.g., the text itself, hashtags, mentions, etc.
Here, two sets of tweets were gathered: (i)~a \textit{baseline} dataset with $1M$ random tweets, and (ii)~a \textit{Gamergate}-related dataset with $650k$ tweets.

\descr{Gamergate dataset.} 
To build a dataset containing an adequate number of bullying / aggressive instances, we initially selected \#GamerGate as a seed word.
From the $1\%$ sample of public tweets, we selected only those containing this seed word and performed a snowball sampling of other hashtags likely associated with abusive behavior.
Thus, we included tweets which contained hashtags that appeared in the same tweets as \#GamerGate (the keywords list was updated on a daily basis - more details about the data collection process can be found in our previous work~\cite{chatzakou2017Gamergate}).
Overall, we collected $308$ hashtags during the data collection period.
After a manual examination of these hashtags, we verified that they indeed contain a number of abusive words or hashtags, e.g., \#InternationalOffendAFeministDay, \#IStandWithHateSpeech, and \#KillAllNiggers.

\descr{Baseline (random) dataset.} 
To compare the hate-related dataset with cases which are less prone to contain abusive content, and for the same time period, we also crawled a random sample of $1M$ tweets which serve as a baseline.

\subsection{Preprocessing}

Next, we focus on the tasks performed to make our data suitable for analysis, cleaning text, and removing noise, and dealing with other erroneous data.

\descr{Cleaning.} We remove stop words, numbers, and punctuation marks.
Also, we normalize text by eliminating repetitive characters which users often use to express their feelings with more intensity (e.g., the word `hellooo' is converted to `hello').
Users tend to add extra vowels in words to show emphasis or intense emotion.
So, based on such an assumption, initially we remove all the duplicate vowels (only when they are above 2) of a word, if any.
Then, we check for the existence of the ``new'' word in the Wikipedia database.
Such process is repeated for all the possible combinations when more than one vowels is duplicate.
If none of the ``new'' words is available in the Wikipedia database, we keep the initial one.

\descr{Spam removal.} Even though extensive work has been done on spam detection in social media, e.g.,~\cite{stringhini2010detecting,Wang2010SpamDetectionTwitter}, Twitter is still plagued by spam accounts~\cite{Chen2015SpamTweets}.
Two main indications of spam behavior are~\cite{Wang2010SpamDetectionTwitter}: (i)~the large number of hashtags within a user's posts, as it permits the broader broadcast of such posts, and (ii)~the population of large amounts of (almost) similar posts.
Based on the 2-month dataset collected from Twitter, the distributions of hashtags and duplications of posts are examined to detect the cutoff-limit above which a user will be characterized as spammer and consequently will be removed from the dataset.

\emph{Hashtags.} Studying the hashtags distribution, we observe that users use on average $0$ to $17$ hashtags.
Building on this, we examine various cuttoffs to select a proper one above which we can characterize a user as spammer.
In the end, after a manual inspection we observed that in most of the cases where the number of hashtags was $5$ or more, the text was mostly related to inappropriate content.
So, the limit of $5$ hashtags is used, and consequently we remove those users that have more than $5$ hashtags on average in their tweets.

\emph{Duplications.} In many cases a user's texts are (almost) the same, with only the listed mentioned users modified.
So, in addition to the previously mentioned cleaning processes, we also remove all mentions.
Then, to estimate the similarity of a user's posts we proceed with the Levenshtein distance~\cite{Navarro2001ApproximateStringMatching} which counts the minimum number of single-character edits needed to convert one string into another, averaging it out over all pairs of their tweets.
Initially, for each user we calculate their intra-tweets similarity.
Thus, for a user with $x$ tweets, we arrive at a set of $n$ similarity scores, where $n = x (x - 1) / 2$, and an average intra-tweet similarity per user.
Then, all users with average intra-tweets similarity above $0.8$ (about $5\%$) are excluded from the dataset.

\section{Comparing Gamergaters with typical users} \label{sec:comparison}

In this section, we compare the baseline and GG-related dataset across two dimensions, considering emotional and activity attributes.

\subsection{Emotional characteristics of Gamergaters}

\begin{figure*}[!t]
	\centering
	\begin{subfigure}[b]{0.24\textwidth}
		\includegraphics[width=\textwidth]{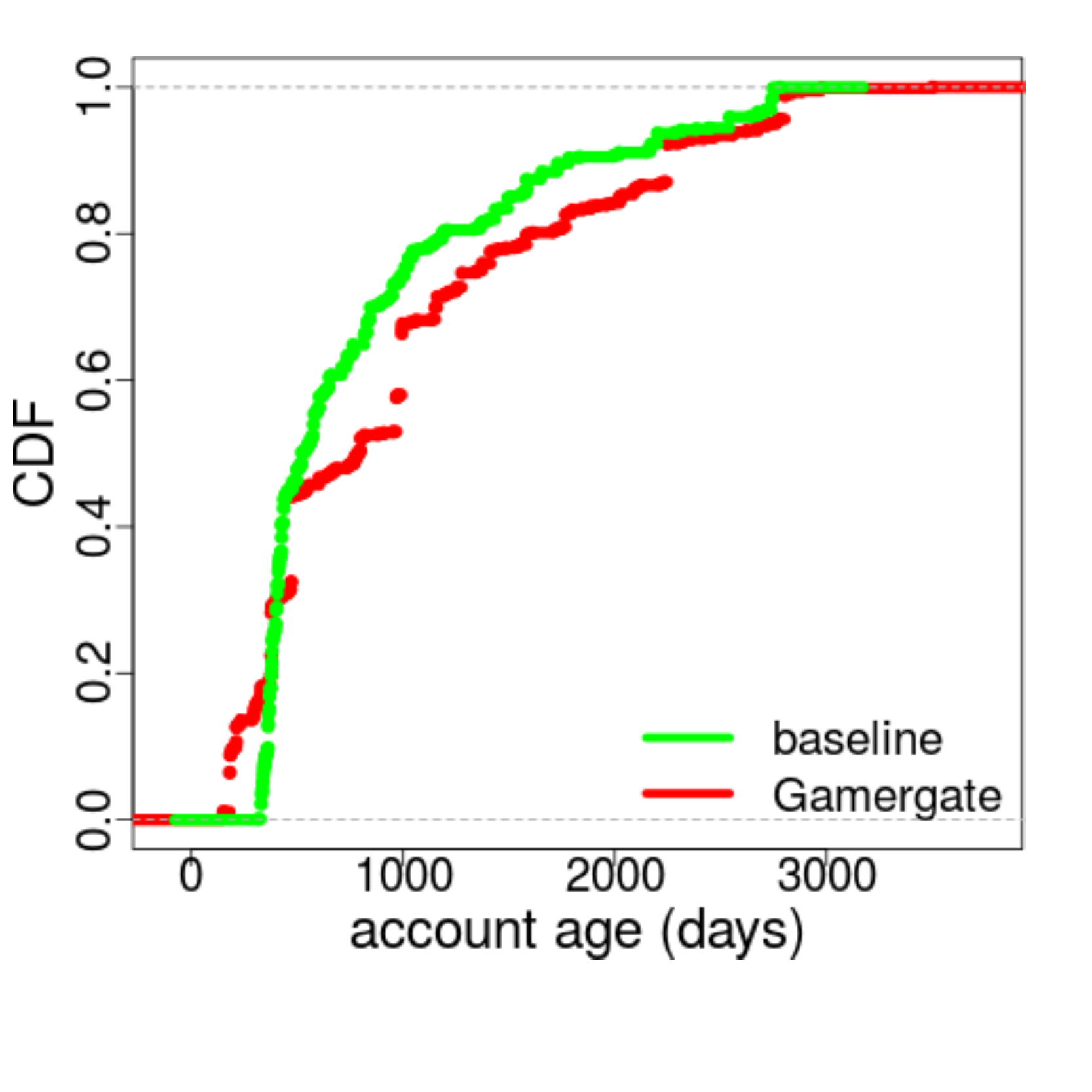}
		\captionsetup{font=scriptsize}
		\vspace{-0.95cm}
		\caption{Account age distribution.}
		\label{fig:baseline_hatebase_age}
	\end{subfigure}
	\begin{subfigure}[b]{0.24\textwidth}
		\includegraphics[width=\textwidth]{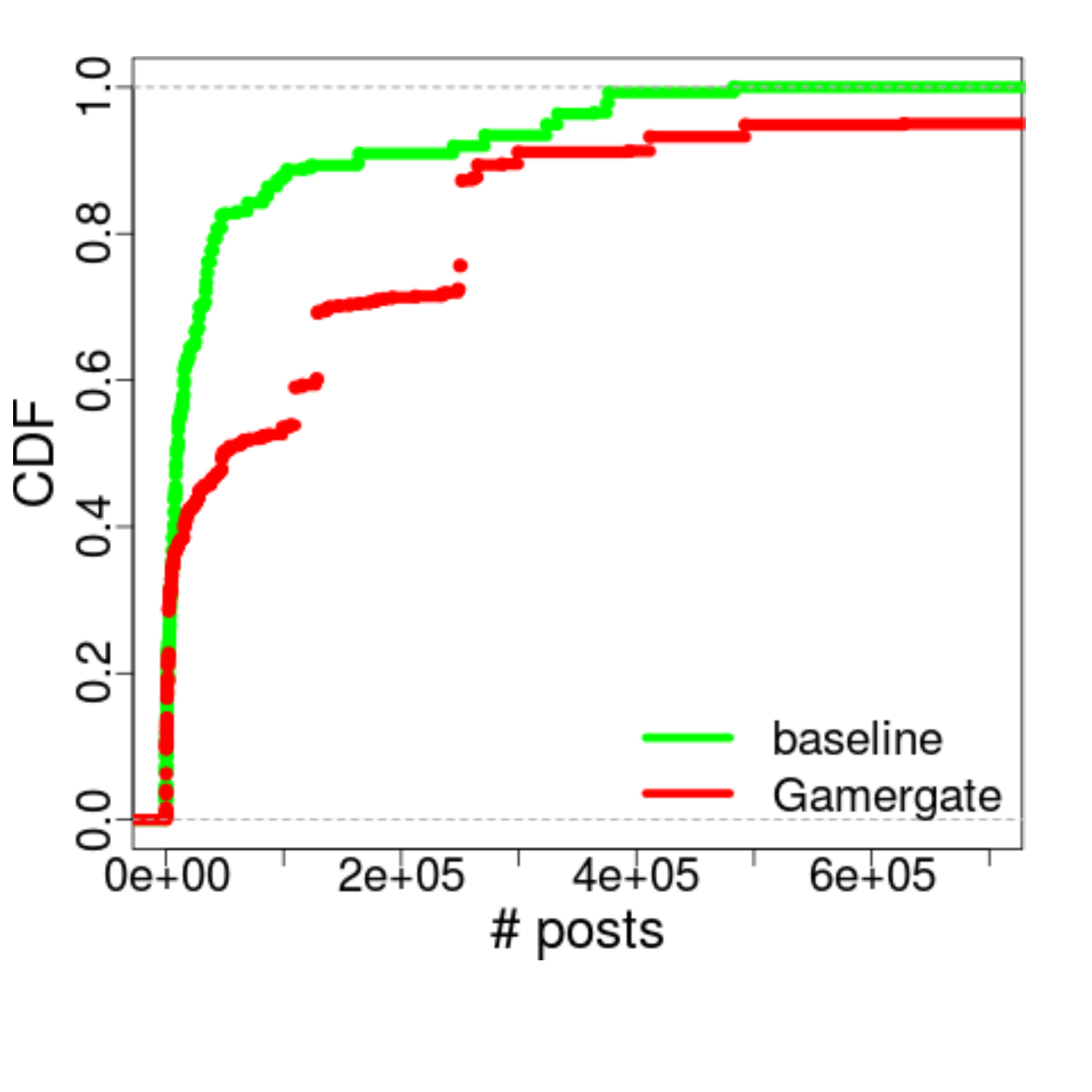}
		\captionsetup{font=scriptsize}
		\vspace{-0.95cm}
		\caption{Number of posts distribution.}
		\label{fig:baseline_hatebase_posts}
	\end{subfigure}
	\begin{subfigure}[b]{0.24\textwidth}
		\includegraphics[width=\textwidth]{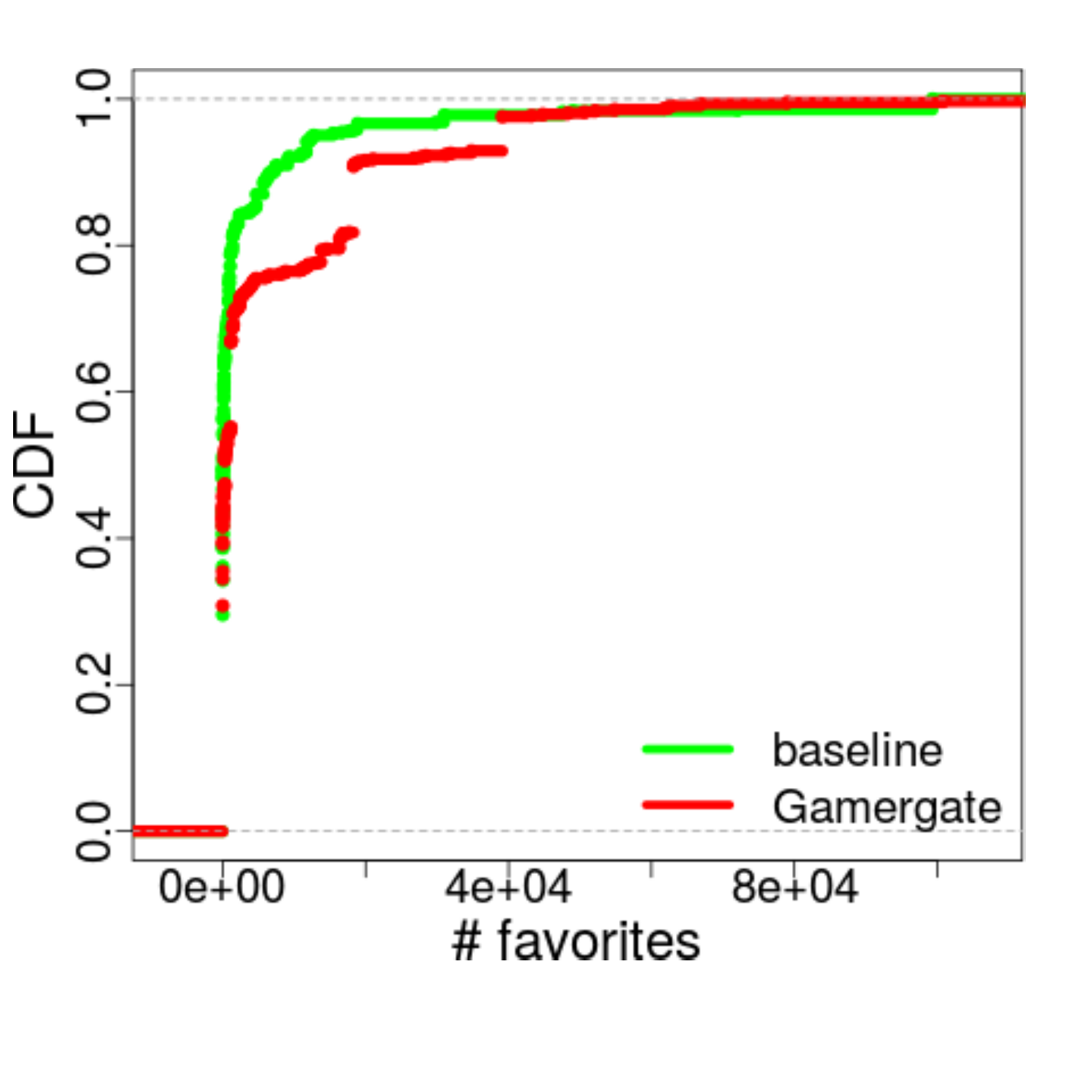}
		\captionsetup{font=scriptsize}
		\vspace{-0.95cm}
		\caption{Favorites distribution.}
		\label{fig:baseline_hatebase_favourites}
	\end{subfigure}
	\begin{subfigure}[b]{0.24\textwidth}
		\includegraphics[width=\textwidth]{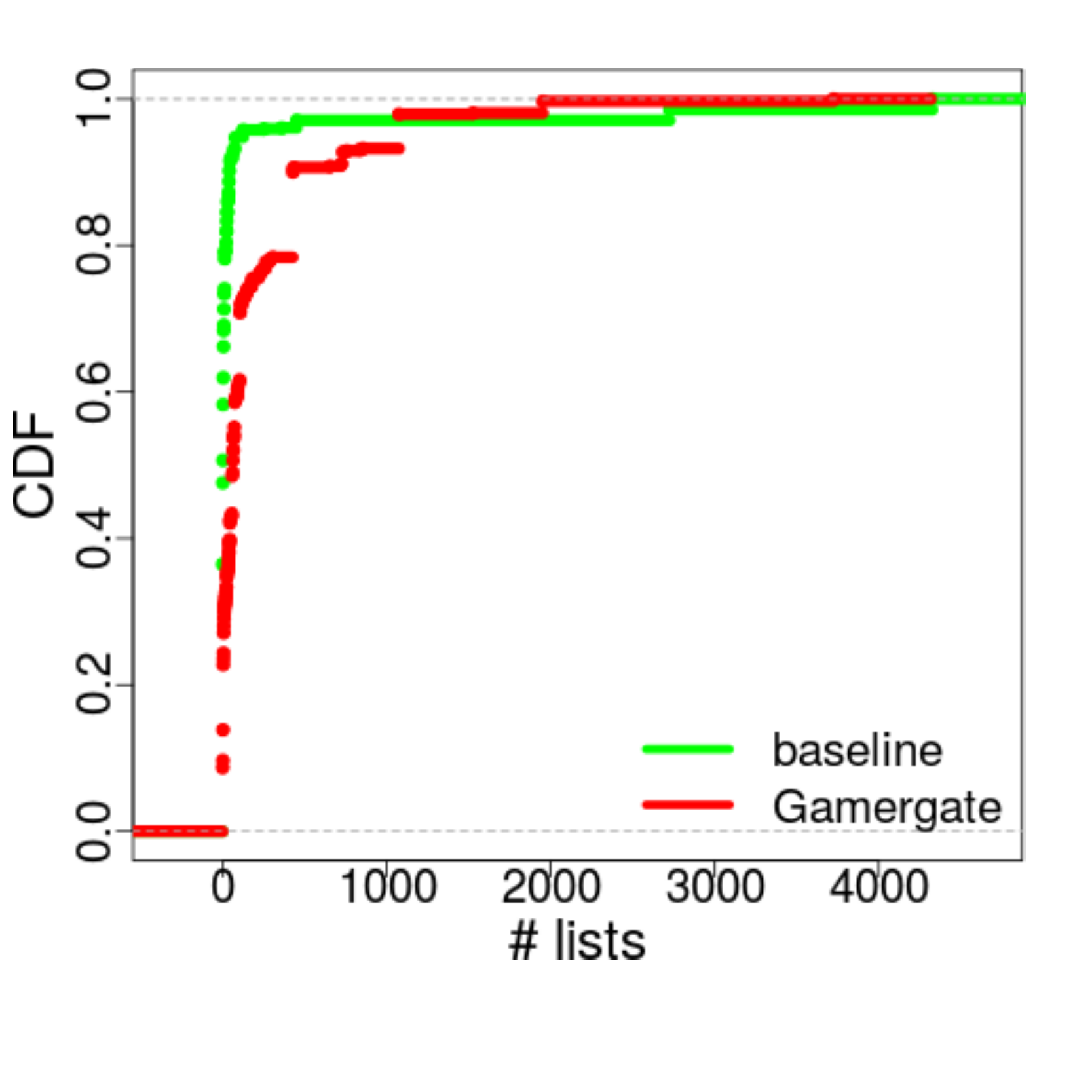}
		\captionsetup{font=scriptsize}
		\vspace{-0.95cm}
		\caption{Lists distribution.}
		\label{fig:baseline_hatebase_lists}
	\end{subfigure}	
	\begin{subfigure}[b]{0.24\textwidth}
		\includegraphics[width=\textwidth]{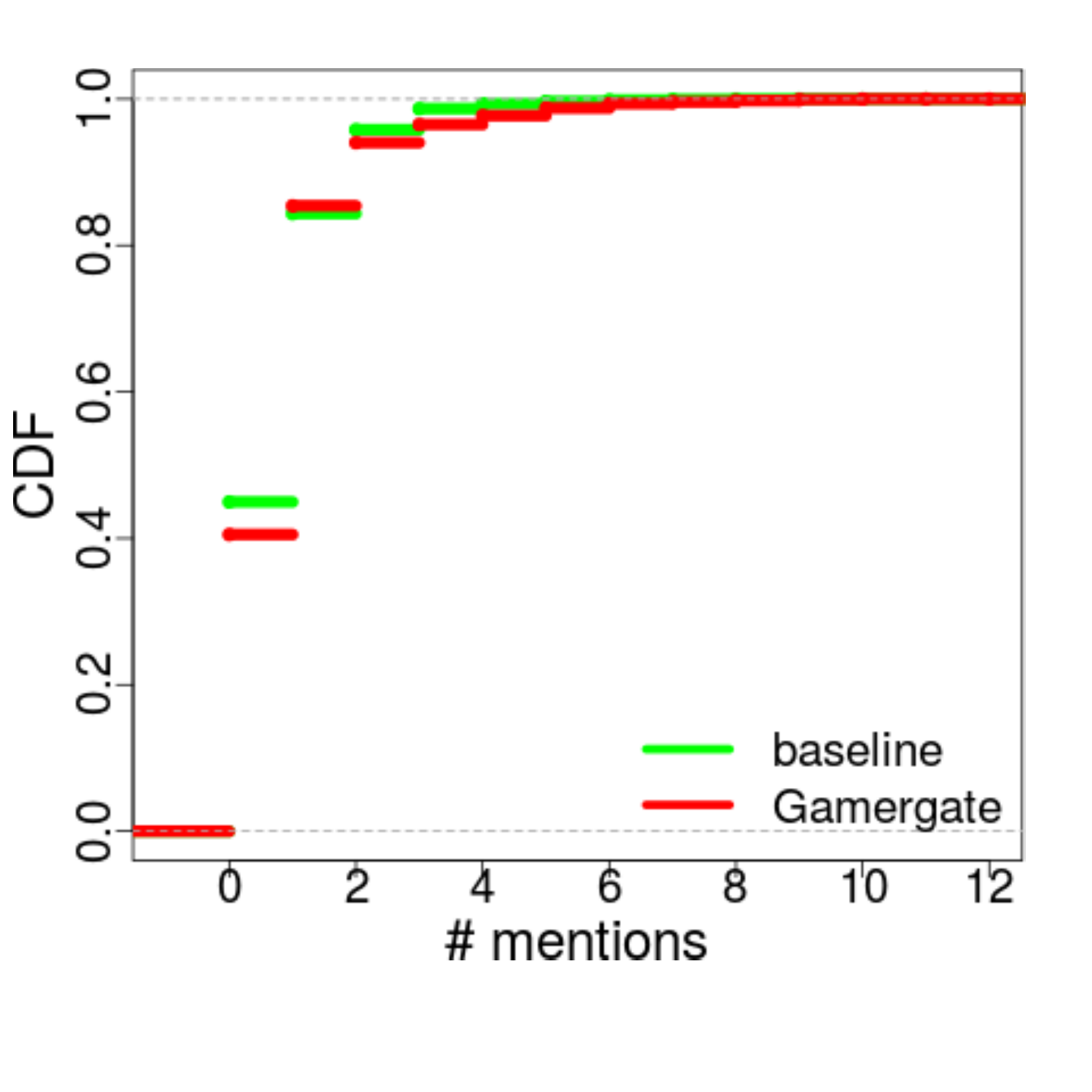}
		\captionsetup{font=scriptsize}
		\vspace{-0.95cm}
		\caption{Mentions distribution.}
		\label{fig:baseline_hatebase_mentions}
	\end{subfigure}	
	\begin{subfigure}[b]{0.24\textwidth}
		\includegraphics[width=\textwidth]{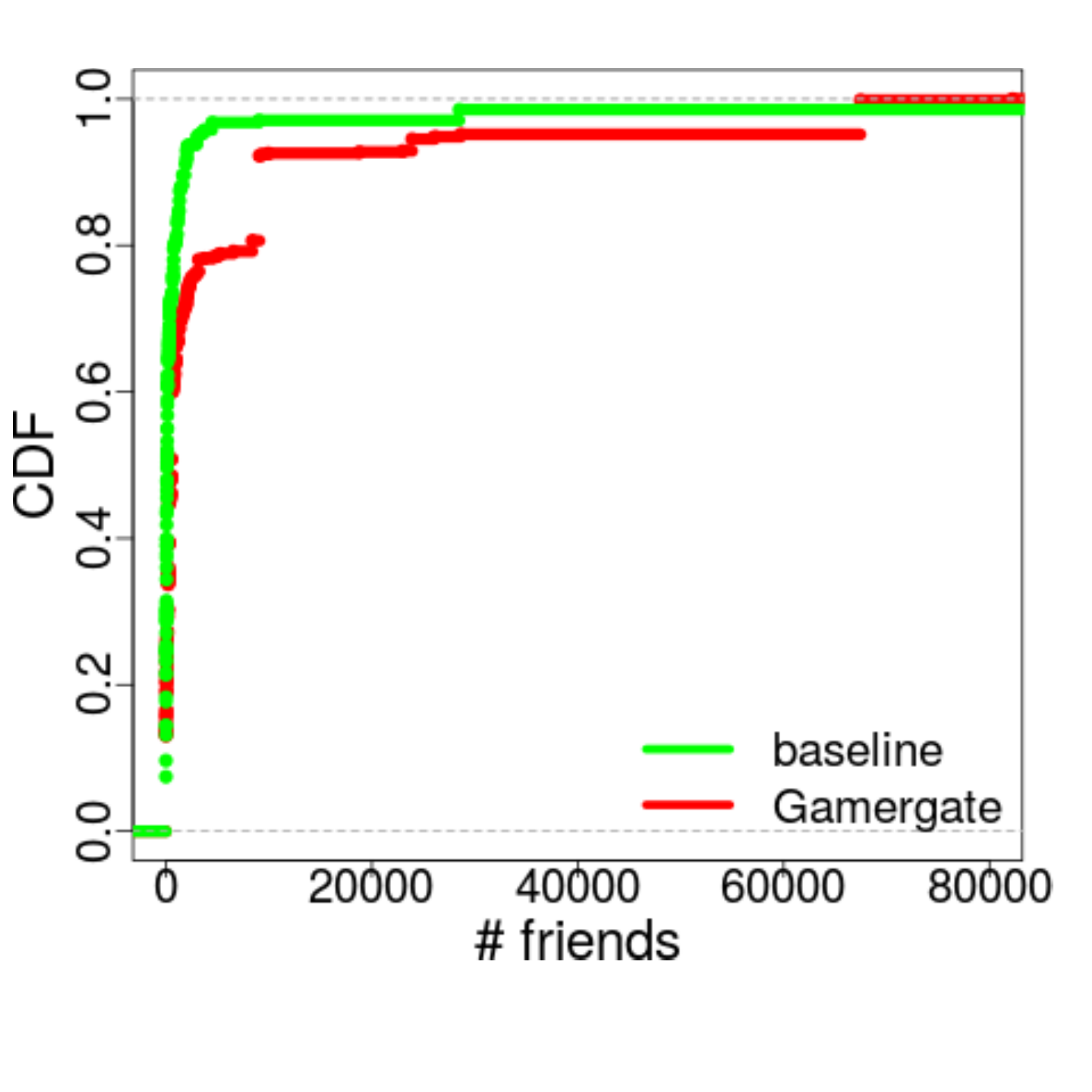}
		\captionsetup{font=scriptsize}
		\vspace{-0.95cm}
		\caption{Friends distribution.}
		\label{fig:baseline_hatebase_friends}
	\end{subfigure}
	\begin{subfigure}[b]{0.24\textwidth}
		\includegraphics[width=\textwidth]{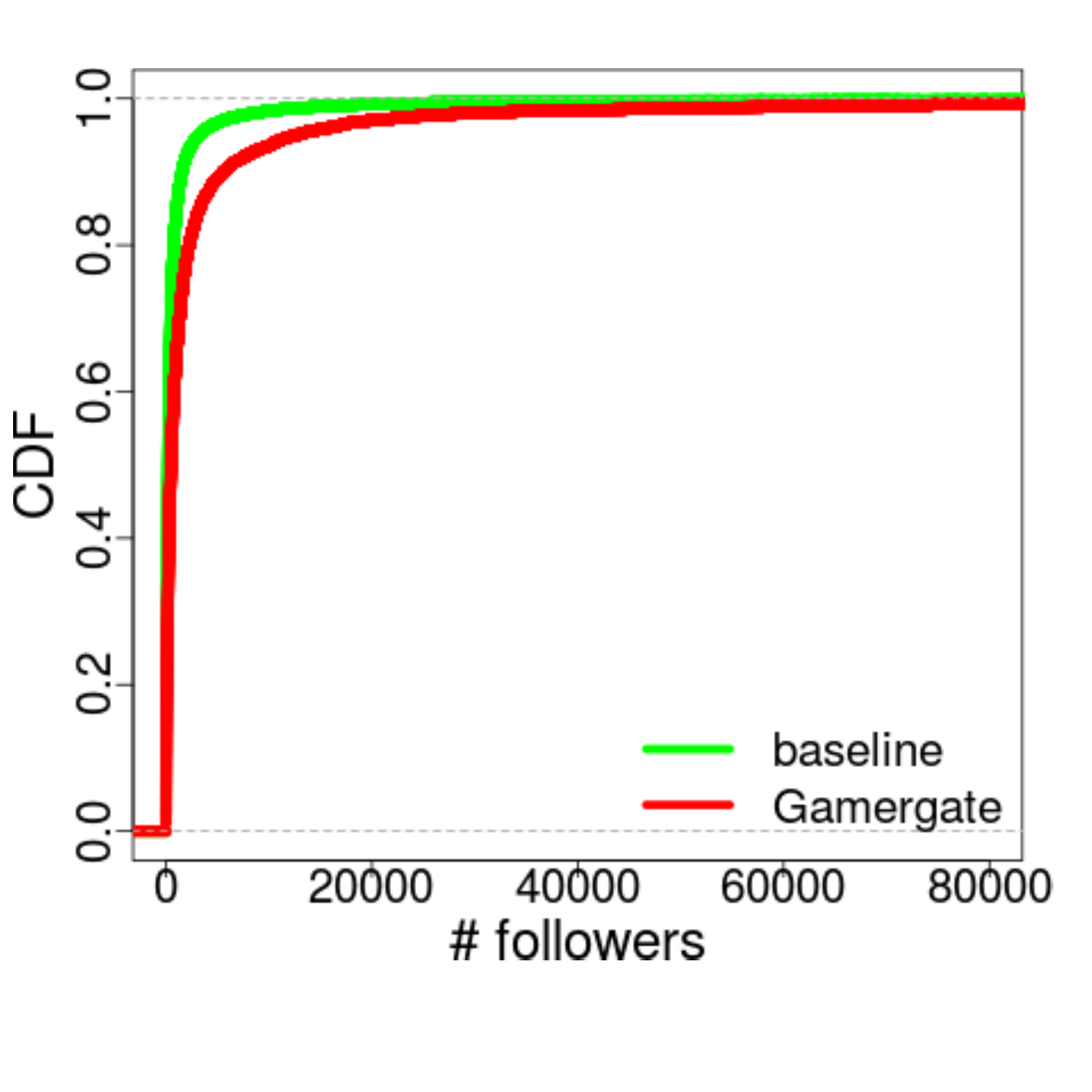}
		\captionsetup{font=scriptsize}
		\vspace{-0.95cm}
		\caption{Followers distribution.}
		\label{fig:baseline_hatebase_followers}
	\end{subfigure}	
	\vspace{-0.15cm}
	\caption{ CDF distribution of (a) Account age, (b) Number of Posts, (c) Favorites, (d) Lists, (e) Mentions, (f) Friends, (g) Followers.}
\end{figure*}

\descr{Sentiment.}
To detect sentiment, we use the SentiStrength tool~\cite{sentistrength}, which estimates the positive and negative sentiment (on a [-4, 4] scale) in short texts.
Figure~\ref{fig:baseline_hatebase_sentiment} plots the CDF of sentiment of tweets for the two datasets.
We note that around $25\%$ of tweets are positive for both types of users.
However, GGers post tweets with a generally more negative sentiment (a two-sample Kolmogorov-Smirnov test rejects the null hypothesis with $D=0.101$, $p < 0.01$).
In particular, around $25\%$ of GG tweets are negative compared to only around $15\%$ for baseline users.
This observation is in line with the GG dataset containing a large number of offensive posts.

\descr{Emotions.}
We also extract the sentiment values for six emotions using a similar approach to~\cite{Chatzakou2013EmotionallyDrivenClustering}: anger, disgust, fear, joy, sadness, and surprise which, based on Ekman et al.~\cite{Ekman1982}, are considered as \emph{primary} emotions.
Also known as basic, they are a fixed number of emotions which we experience instantly as a response to a pleasant (or unpleasant) stimulus.
Figure~\ref{fig:baseline_hatebase_joy} shows the CDF of joy, where we reject the null hypothesis that the two distributions are the same ($D=0.089$, $p < 0.01$).
We are \emph{unable} to reject the null hypothesis for the other five primary emotions.
This is particularly interesting because it contradicts the narrative that GGers are posting virulent content out of anger.
Instead, GGers appear to be less joyful.
This is a subtle but important difference: GGers are not necessarily angry, but they are apparently less happy.

\descr{Offensive.}
Looking a bit deeper, we compare the offensiveness score that tweets have been marked with according to the hatebase (HB)~\cite{hatebase} crowdsourced dictionary.
Each word included in HB is scored on a [0, 100] scale which indicates how hateful it is.
Though the visual difference is small, GGers use more hateful words than a baseline user ($D=0.006$, $p < 0.01$).

\descr{Emoticons and Uppercase.}
Two common ways to express emotion in social media are emoticons and ``shouting'' by using all capital letters.
Based on the nature of GG, we initially suspected that there would be a relatively small amount of emoticon usage, but many tweets that would be shouting in all uppercase letters.
However, as we can see in Figures~\ref{fig:baseline_hatebase_emoticons} and~\ref{fig:baseline_hatebase_uppercases}, which plot the CDF of the average number (per user) of emoticon usage and all uppercase tweets, respectively, this is not the case.
GG and baseline users tend to use emoticons similarly (we are unable to reject the null hypothesis with $D=0.028$ and $p = 0.96$).
However, GGers tend to use all uppercase \emph{less} than baseline users ($D=0.212$, $p < 0.01$).
As seen previously, GGers are quite savvy Twitter users, and generally speaking, shouting tends to be ignored.
Thus, one explanation is that GGers avoid such a simple ``tell'' as posting in all uppercase to ensure their message is not so easily dismissed.

\subsection{Activity characteristics of Gamergaters}

\descr{Account age.}
An underlying question about GG is what started first: participants' use of Twitter or their participation in the controversy.
I.e., did Gamergate draw people to Twitter, or were Twitter users drawn to Gamergate?
Figure~\ref{fig:baseline_hatebase_age} plots the distribution of account age for GG participants and baseline Twitter users.
For the most part, GGers tend to have older accounts than baseline Twitter users ($D = 0.20142$, $p < 0.01$, $mean = 982.94$ days, $median = 788$ days, $STD = 772.49$ days).
The mean, median, and STD values for the baseline users are $834.39$, $522$, and $652.42$ days, respectively.
Overall, the oldest account in our dataset belongs to a GG user, while only $26.64\%$ of baseline users have account ages older than the mean value of the GGers.
The figure indicates that GG users were existing Twitter users that were drawn to the controversy.
In fact, their familiarity with Twitter could be the reason that GG exploded in the first place.

\descr{Posts, Favorites, and Lists.}
Figure~\ref{fig:baseline_hatebase_posts} plots the distribution of the number of tweets made by GGers and baseline users.
GGers are significantly more active than baseline Twitter users ($D = 0.352$, $p < 0.01$).
The mean, median and STD values for the GG (random) users is $135,618$ ($49,342$), $48,587$ ($9,429$), and $185,997$ ($97,457$) posts, respectively.
Figures~\ref{fig:baseline_hatebase_favourites} and~\ref{fig:baseline_hatebase_lists} show the CDFs of favorites and lists declared in users' profiles.
We note that in the median case, GGers are similar to baseline users, but looking at the $30\%$ of users in the tail of each distribution, GG users have more favorites and lists than baseline users.

\descr{Mentions.}
Figure~\ref{fig:baseline_hatebase_mentions} shows that GGers tend to make more mentions within their posts, which can be due to the higher number of direct attacks in contrast to the baseline users.

\descr{Followers and Friends.}
GGers are involved in what we would typically think of as anti-social behavior.
However, this is somewhat at odds with the fact that their activity takes place primarily on social media.
To get an idea of how ``social'' GGers are, Figures~\ref{fig:baseline_hatebase_friends} and~\ref{fig:baseline_hatebase_followers} plot the distribution of friends and followers for GGers and baseline users.
We observe that GGers tend to have more friends and followers than baseline twitter users ($D=0.34$ and $0.39$, $p < 0.01$ for both).
Although this result might be initially counter-intuitive, the truth of the matter is that GG was born on social media, and is a very clear ``us vs. them'' situation.
This leads to easy identification of in-group membership, and thus heightens the likelihood of relationship formation.

\section{Suspension of Gamergate Accounts by Twitter} \label{sec:statuses}

In the previous section, we studied users involved in the GG controversy and identified attributes that distinguish them from random Twitter users, either regarding the way they write tweets and the sentiment they carry, or their embeddedness in the Twitter social network.
In fact, we found that GGers post tweets that are more negative, less joyful, and more hateful or offensive.
However, we also observed that such users have more friends and followers, more posting and dissemination activity (via hashtags and mentions).
From this clearly distinctive behavior, what remains unclear is how these users are handled by Twitter.

To shed more light on this aspect, in the next sections, we examine further the GGers by introducing a new factor characterizing each one: their Twitter account status.
In particular, we investigate the following questions:
\begin{squishlist}
\item What is the twitter account status and how do we measure it? What does it imply for a user and what is the breakdown for different statuses between GGers and random users (\S~\ref{subsec:twitter-status})?
\item What are the characteristics of suspended users and users who deleted their Twitter account (\S~\ref{subsec:suspended-deleted})?
\item What are the characteristics of users who remain active on Twitter, but should have been suspended (\S~\ref{subsec:should-suspend})?
\item Can we emulate the Twitter account suspension mechanism (\S~\ref{subsec:status-classification})?
\end{squishlist}

\descr{Methodology.}
To answer these questions, we analyze users on features presented in the previous section, under the following two general categories:
\begin{squishlist}
\item \emph{emotional attributes}: sentiment, 6 emotions (anger, disgust, fear, joy, sadness, surprise), offensive words, uppercases, emoticons;
\item \emph{activity attributes}: account age, number of posts, user participating lists, mentions, followers and friends count.
\end{squishlist}
We apply unsupervised and supervised methods to validate that these features are useful to study and compare their distributions to identify differences between types of users and account statuses.

\subsection{Status of Gamergate Accounts on Twitter}
\label{subsec:twitter-status}

A Twitter user can be in one of the following three statuses: (i)~\emph{active}, (ii)~\emph{deleted}, or (iii)~\emph{suspended}.
Typically, Twitter suspends an account (temporarily or even permanently, in some cases) if it has been hijacked/compromised, is considered spam/fake, or if it is \emph{abusive}.\footnote{\url{https://support.twitter.com/articles/15790}}
A user account is \emph{deleted} if the user himself, for his own personal reasons, deactivates his account.

In order to examine the differences between these three statuses in relation to the GGers and baseline users, we selected a $10\%$ random sample of $33k$ users from both the GG ($5k$) and baseline ($28k$) users to check their Twitter status, one month after the initial data collection.
The status of each user's account was checked using a mechanism that queried the Twitter API for each user, and examined the error code responses returned: code $63$ corresponds to a suspended user account and code $50$ corresponds to a deleted one.

\begin{figure}[!t]
	\centering
	\includegraphics[width=0.34\textwidth]{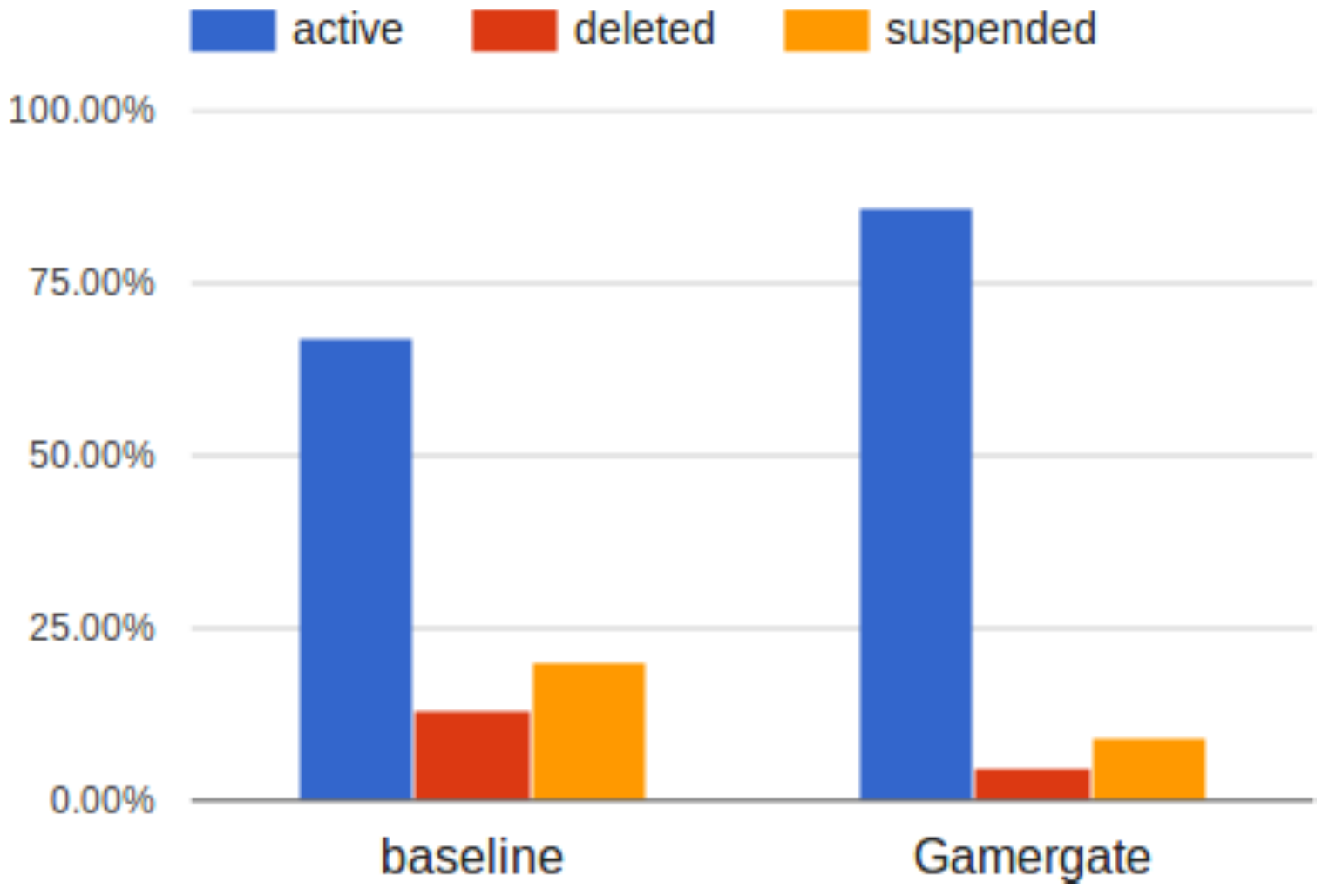}
	\vspace{-0.15cm}
	\caption{Distribution of baseline and GG users in Twitter statuses.}
	\label{fig:status}
	\vspace{-0.4cm}
\end{figure}

From Figure~\ref{fig:status} we observe that both categories of users tend to be suspended rather than deleting their accounts by choice.
However, baseline users are more prone to suspension ($20\%$) and deletion ($13\%$) of their accounts, in contrast to the GGers ($9\%$ and $5\%$, respectively).
The higher number of the suspended and deleted accounts of the baseline users in comparison to GGers is in accordance with the behavior observed in Figure~\ref{fig:baseline_hatebase_age} which shows that the GGers have been in the platform for a longer period than baseline users, meaning they \emph{appear} to be more compliant to Twitter rules.

Nevertheless, this disproportional rate of suspensions for random users with respect to GGers remains a surprising find.
Given our previous observations on their posting behavior, it is unexpected that several of such users are allowed to continue posting tweets.
Indeed, a small portion of these users may be spammers who are difficult to detect and filter out.
That said, Twitter has made significant efforts in addressing spam accounts and we suspect there is a higher presence of such accounts in the baseline dataset, since the GG dataset is more hyper-focused around a somewhat niche topic.

These efforts are less apparent when it comes to the bullying and aggressive behavior phenomena observed on Twitter in general, e.g.,~\cite{salon,guardiantrolls}, and in our present study of GG users, in particular.
However, recently, Twitter has increased its efforts to combat the existing harassment cases, for instance, by preventing suspended users from creating new accounts~\cite{CNNtech}, or temporarily limiting users for abusive behavior~\cite{independent}.
Such efforts constitute initial steps to deal with the ongoing war among the abusers, their victims, and online bystanders.
Next, we further analyze the available data to identify metrics that can provide explanations for understanding the Twitter suspension mechanism.

\begin{figure*}[!ht]
	\centering
	\begin{subfigure}[b]{0.23\textwidth}
		\includegraphics[width=\textwidth]{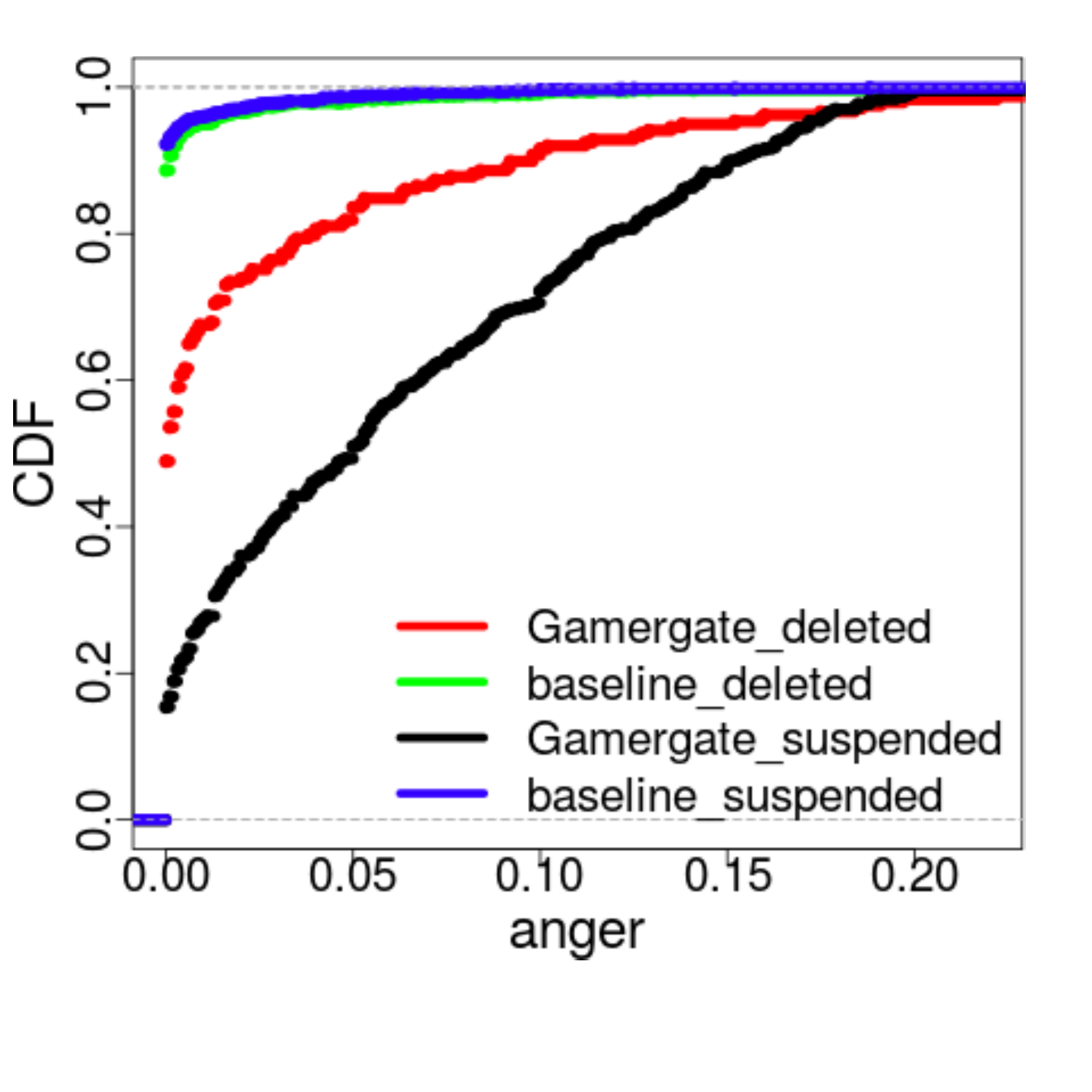}
		\captionsetup{font=scriptsize}
		\vspace{-0.95cm}
		\caption{Anger distribution.}
		\label{fig:sus_del_anger}
	\end{subfigure}
	\begin{subfigure}[b]{0.23\textwidth}
		\includegraphics[width=\textwidth]{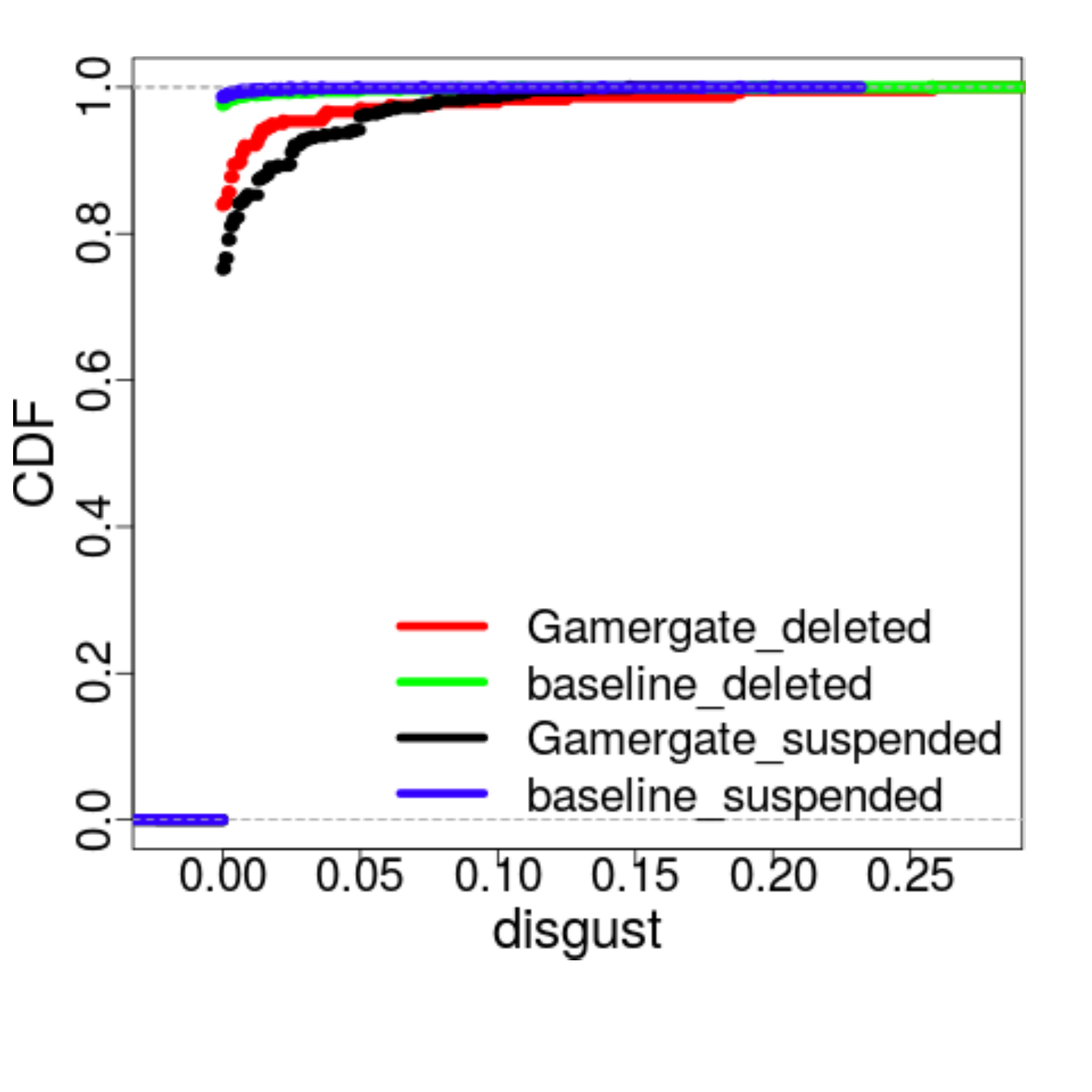}
		\captionsetup{font=scriptsize}
		\vspace{-0.95cm}
		\caption{Disgust distribution.}
		\label{fig:sus_del_disgust}
	\end{subfigure}
	\begin{subfigure}[b]{0.23\textwidth}
		\includegraphics[width=\textwidth]{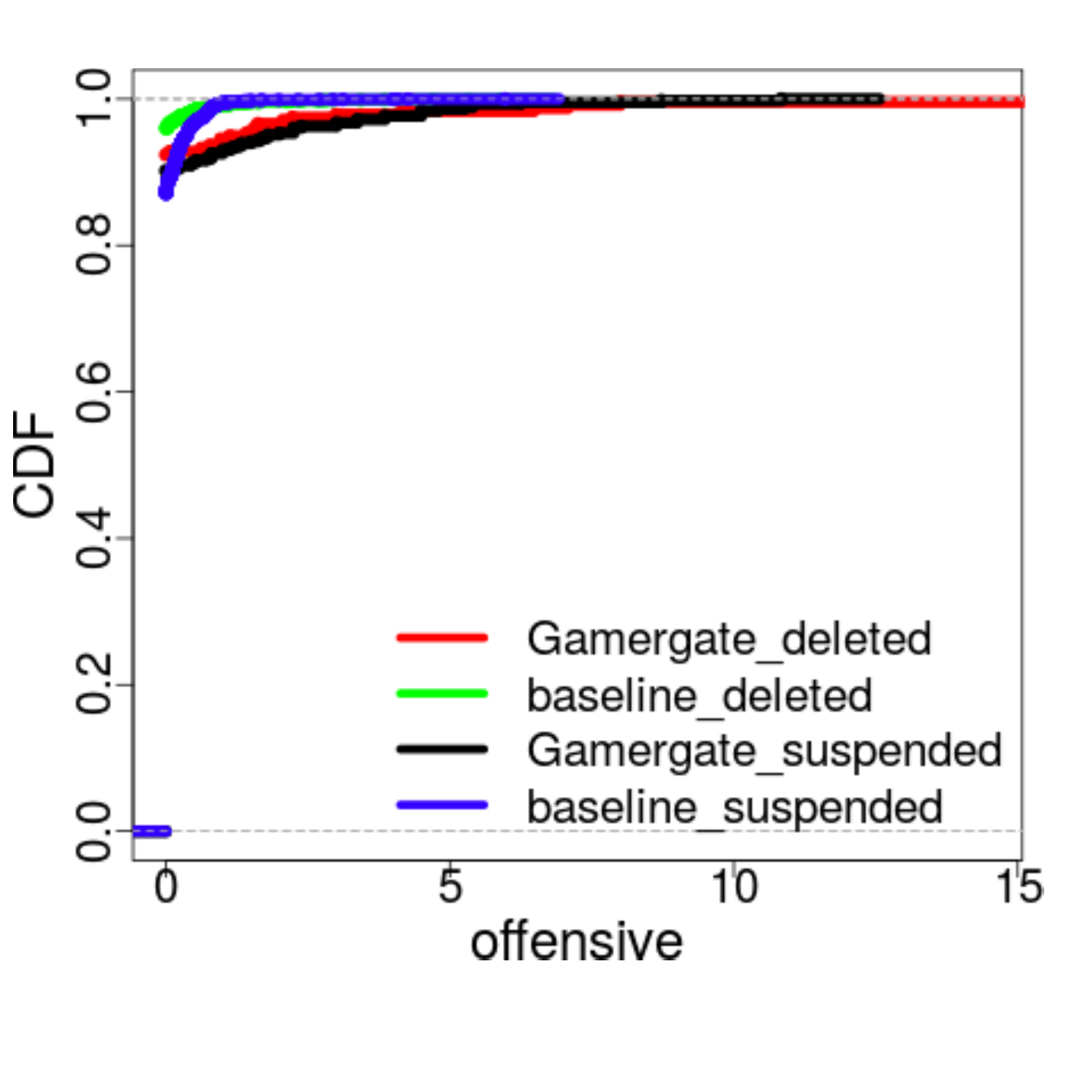}
		\captionsetup{font=scriptsize}
		\vspace{-0.95cm}
		\caption{Offensive distribution.}
		\label{fig:sus_del_offensive}
	\end{subfigure}	
	\begin{subfigure}[b]{0.23\textwidth}
		\includegraphics[width=\textwidth]{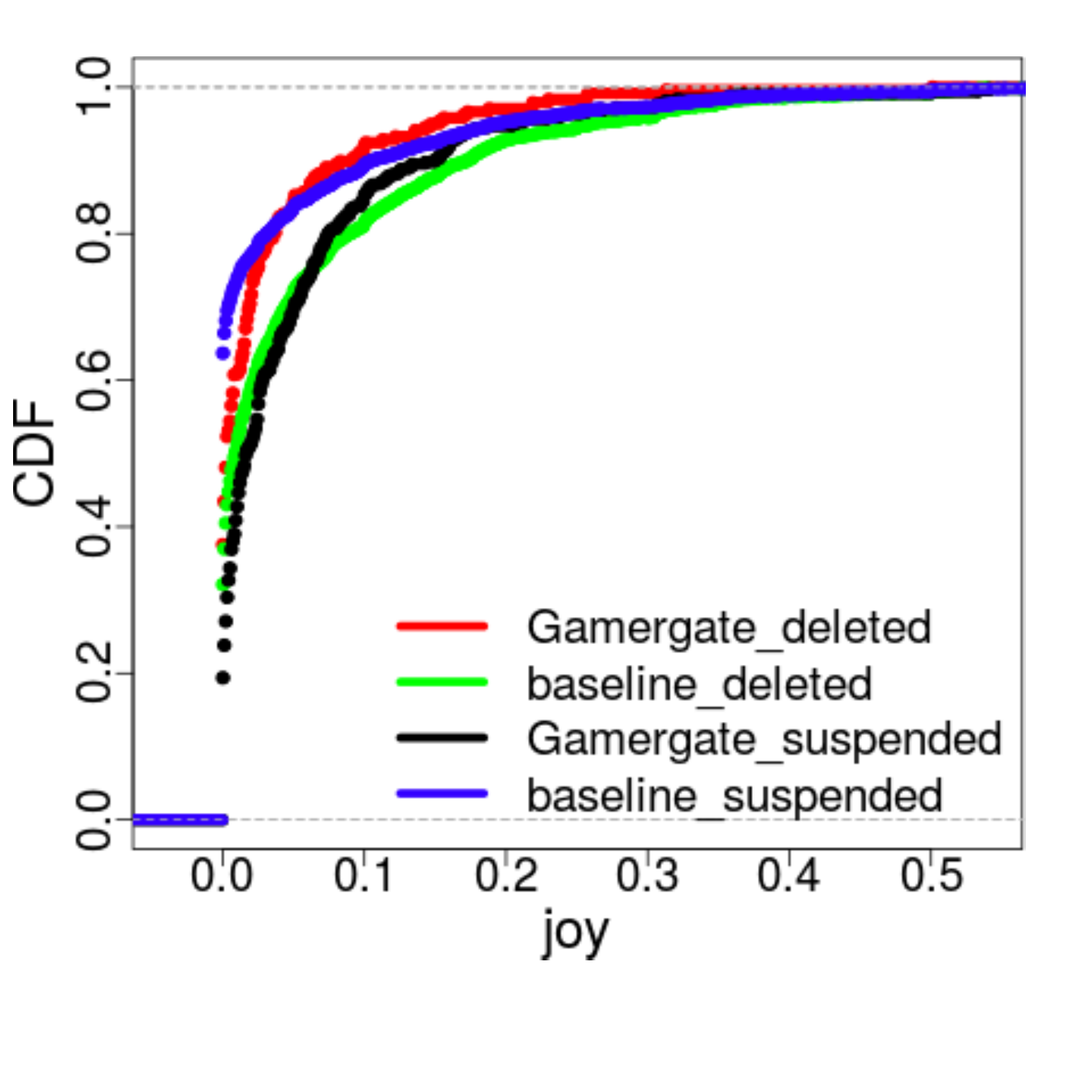}
		\captionsetup{font=scriptsize}
		\vspace{-0.95cm}
		\caption{Joy distribution.}
		\label{fig:sus_del_joy}
	\end{subfigure}
	\begin{subfigure}[b]{0.23\textwidth}
		\includegraphics[width=\textwidth]{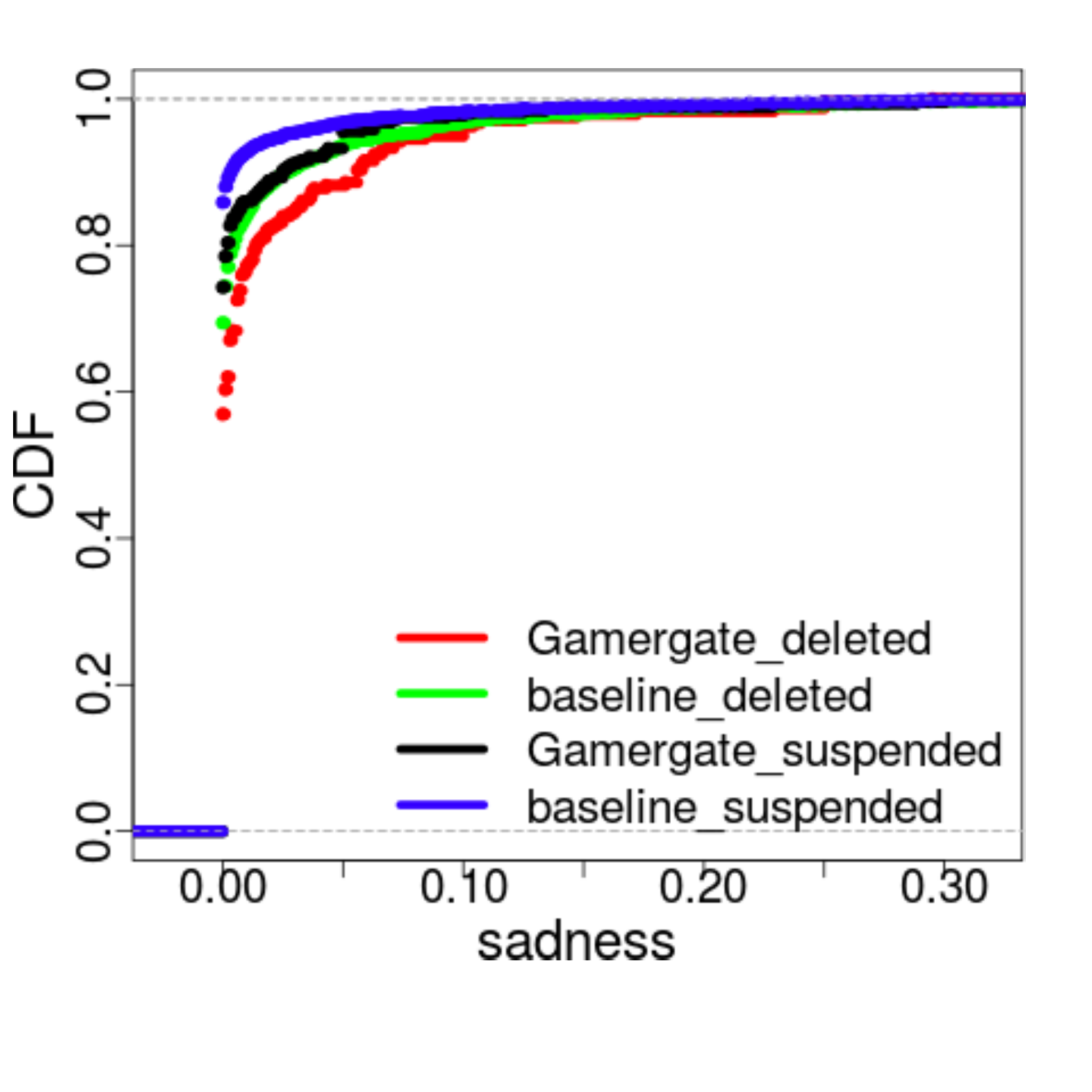}
		\captionsetup{font=scriptsize}
		\vspace{-0.95cm}
		\caption{Sadness distribution.}
		\label{fig:sus_del_sadness}
	\end{subfigure}
	\begin{subfigure}[b]{0.23\textwidth}
		\includegraphics[width=\textwidth]{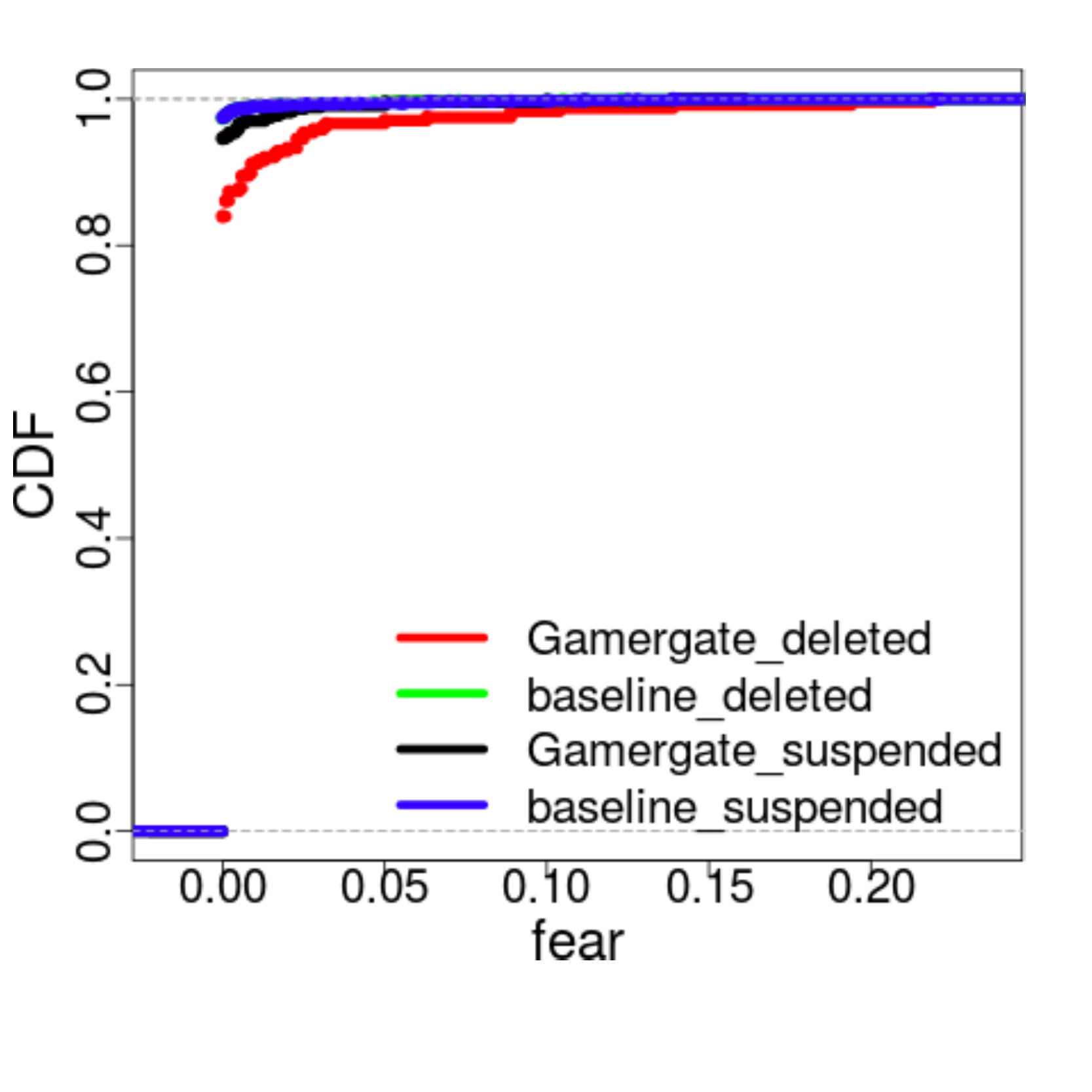}
		\captionsetup{font=scriptsize}
		\vspace{-0.95cm}
		\caption{Fear distribution.}
		\label{fig:sus_del_fear}
	\end{subfigure}	
	\begin{subfigure}[b]{0.23\textwidth}
		\includegraphics[width=\textwidth]{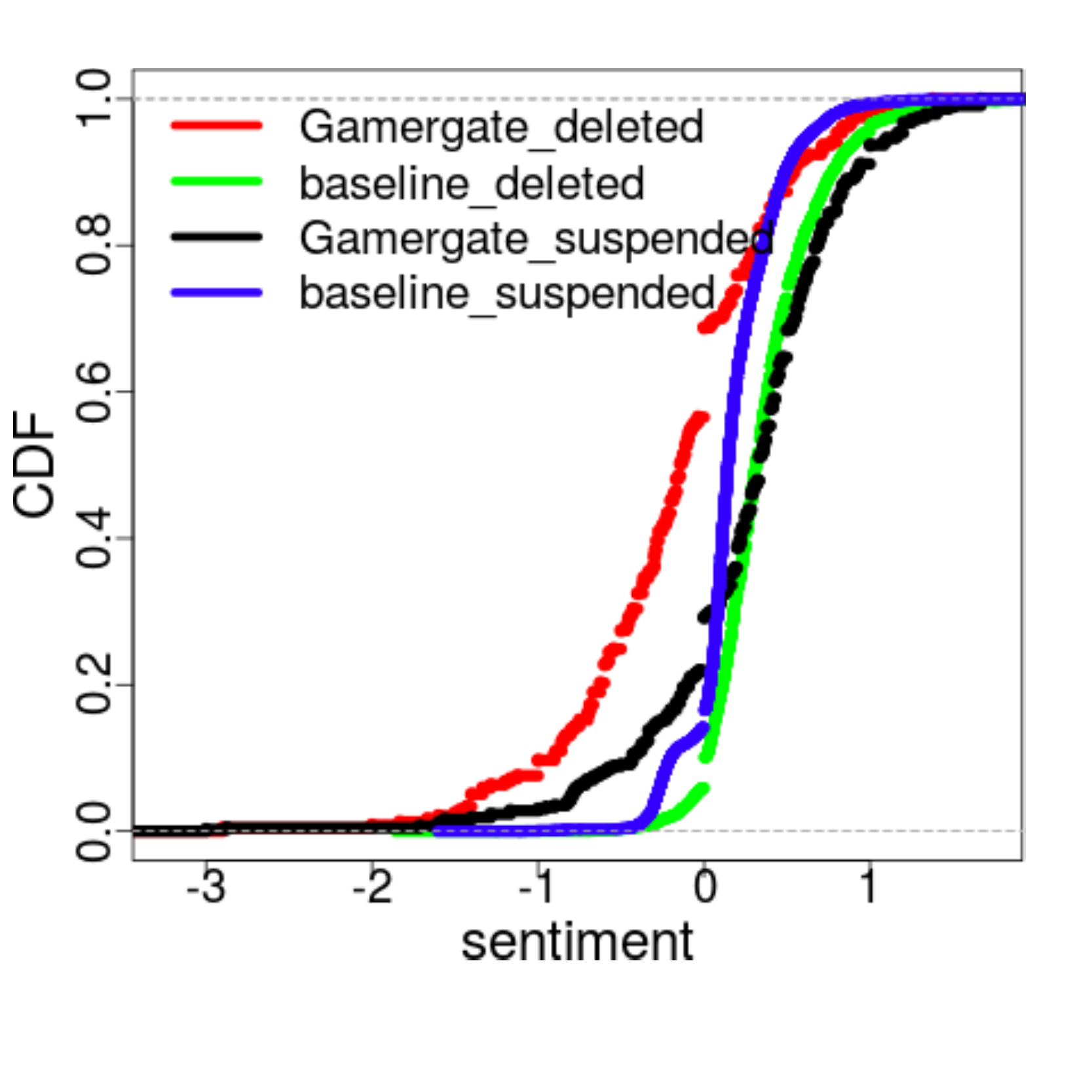}
		\captionsetup{font=scriptsize}
		\vspace{-0.95cm}
		\caption{Sentiment distribution.}
		\label{fig:sus_del_sentiment}
	\end{subfigure}
	\begin{subfigure}[b]{0.23\textwidth}
		\includegraphics[width=\textwidth]{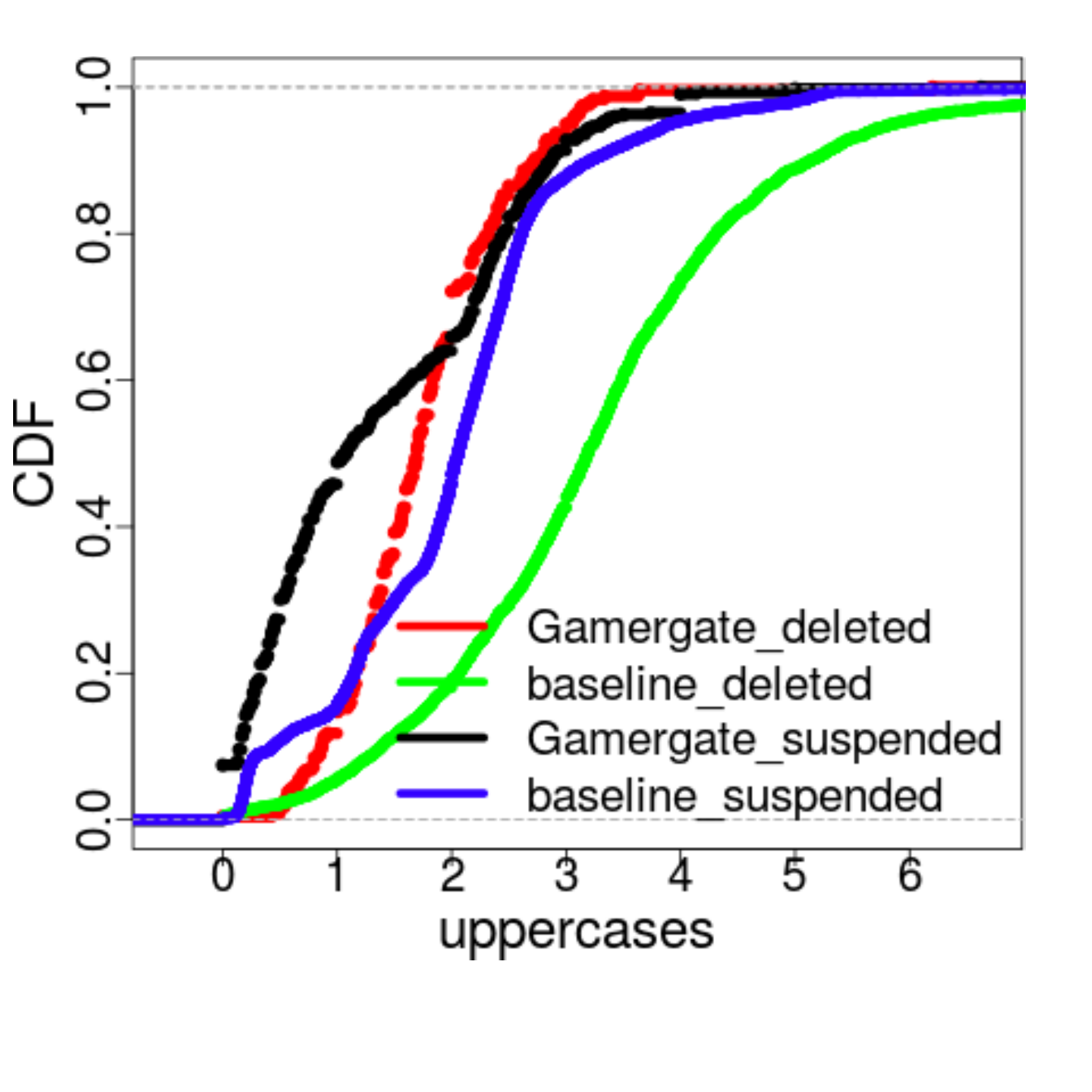}
		\captionsetup{font=scriptsize}
		\vspace{-0.95cm}
		\caption{Uppercases distribution.}
		\label{fig:sus_del_uppercases}
	\end{subfigure}
	\vspace{-0.15cm}
	\caption{CDF plots for the suspended and deleted users considering the emotional attributes: (a) Anger, (b) Disgust, (c) Offensive, (d) Joy, (e) Sadness, (f) Fear, (g) Sentiment, (h) Uppercases.}
	\label{fig:sus_del_emotional}
	\vspace{-0.2cm}
\end{figure*}

\subsection{Who is suspended and who is deleted?}
\label{subsec:suspended-deleted}
To understand how suspended and deleted users differ, here we compare each of these user statuses for both GG and baseline users considering the previously described dimensions, i.e., their emotional and activity based profiles.

\begin{figure}[!t]
	\centering
	\begin{subfigure}[b]{0.23\textwidth}
		\includegraphics[width=\textwidth]{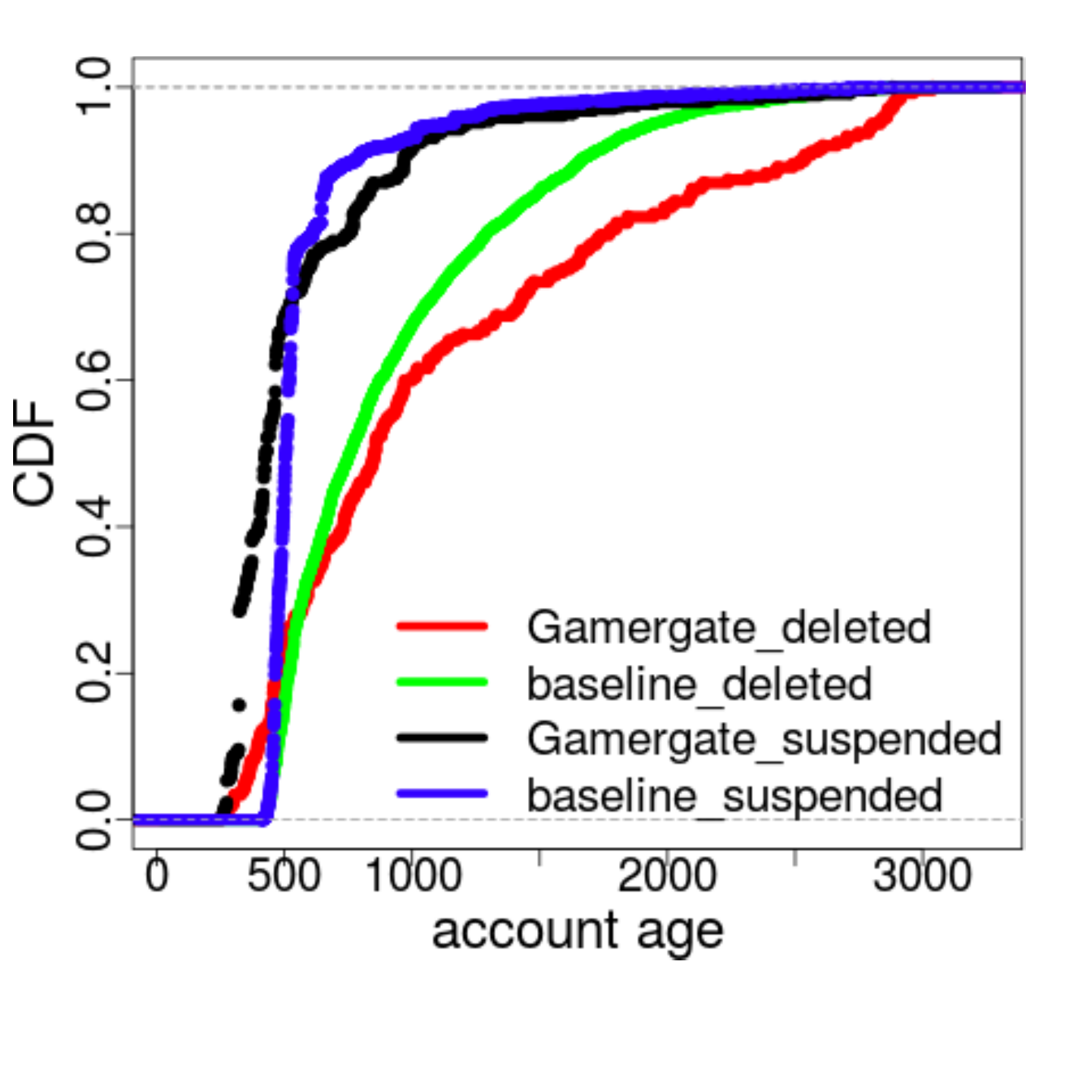}
		\captionsetup{font=scriptsize}
		\vspace{-0.95cm}
		\caption{Account age distribution.}
		\label{fig:sus_del_age}
	\end{subfigure}
	\begin{subfigure}[b]{0.23\textwidth}
		\includegraphics[width=\textwidth]{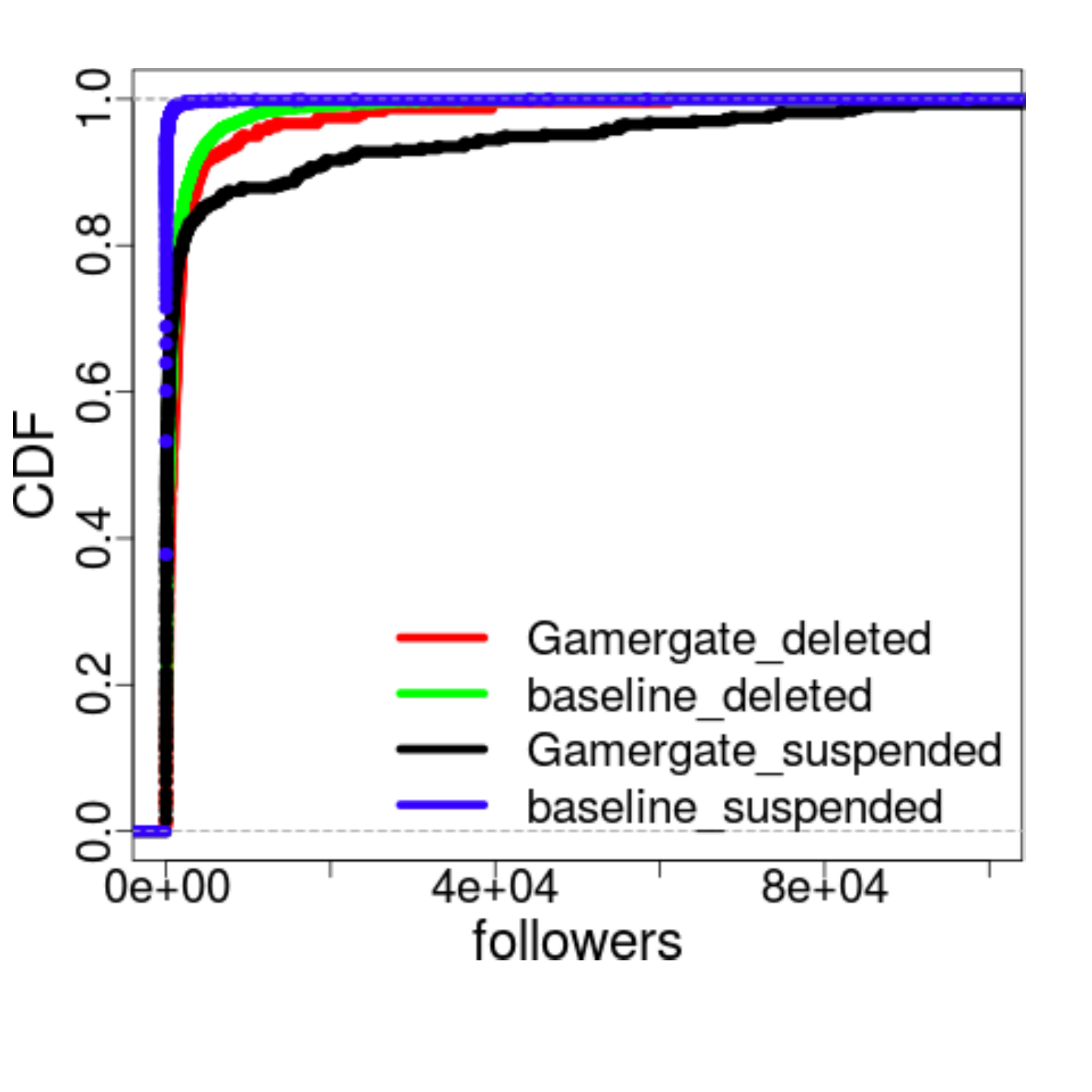}
		\captionsetup{font=scriptsize}
		\vspace{-0.95cm}
		\caption{Followers distribution.}
		\label{fig:sus_del_followers}
	\end{subfigure}
	\begin{subfigure}[b]{0.23\textwidth}
		\includegraphics[width=\textwidth]{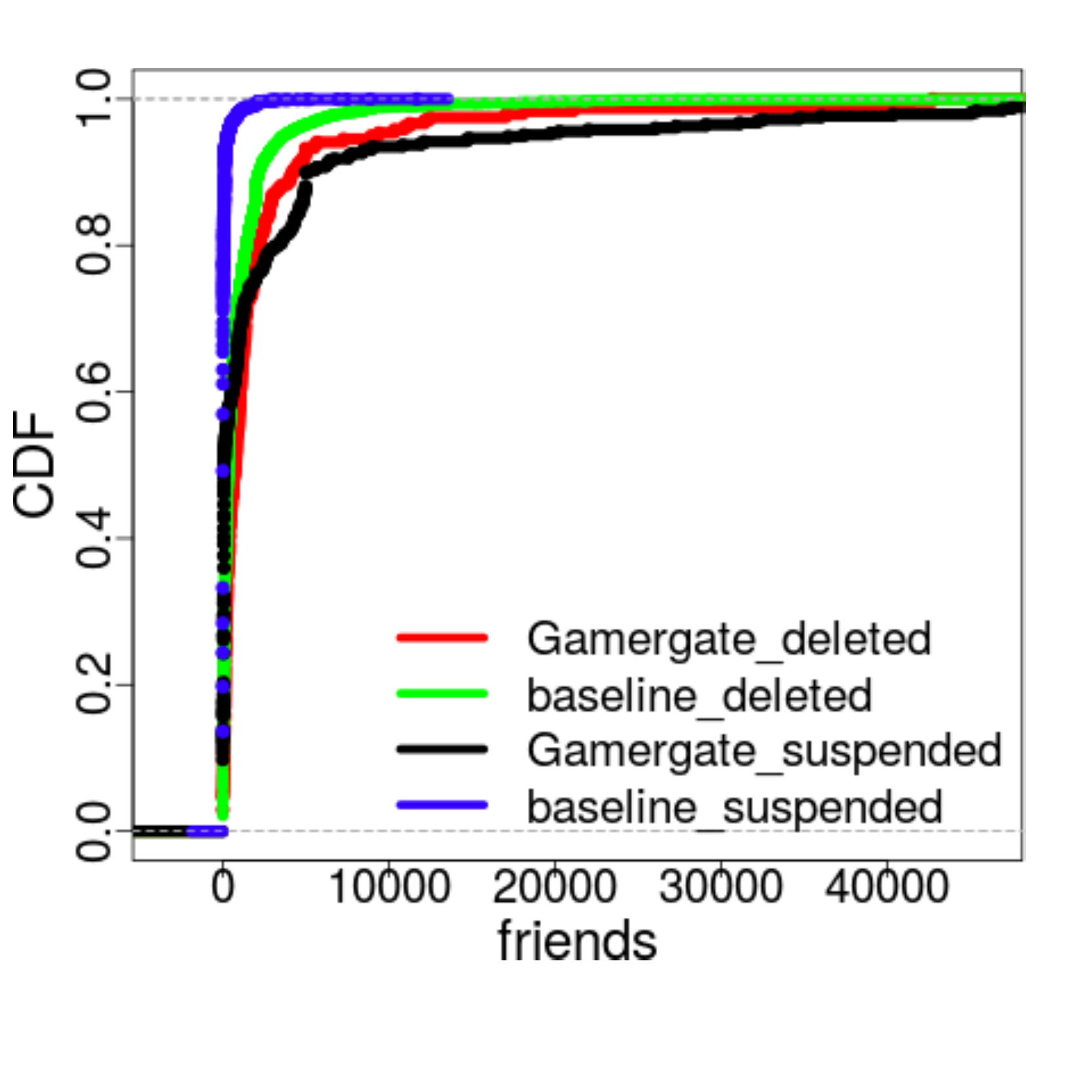}
		\captionsetup{font=scriptsize}
		\vspace{-0.95cm}
		\caption{Friends distribution.}
		\label{fig:sus_del_friends}
	\end{subfigure}	
	\begin{subfigure}[b]{0.23\textwidth}
		\includegraphics[width=\textwidth]{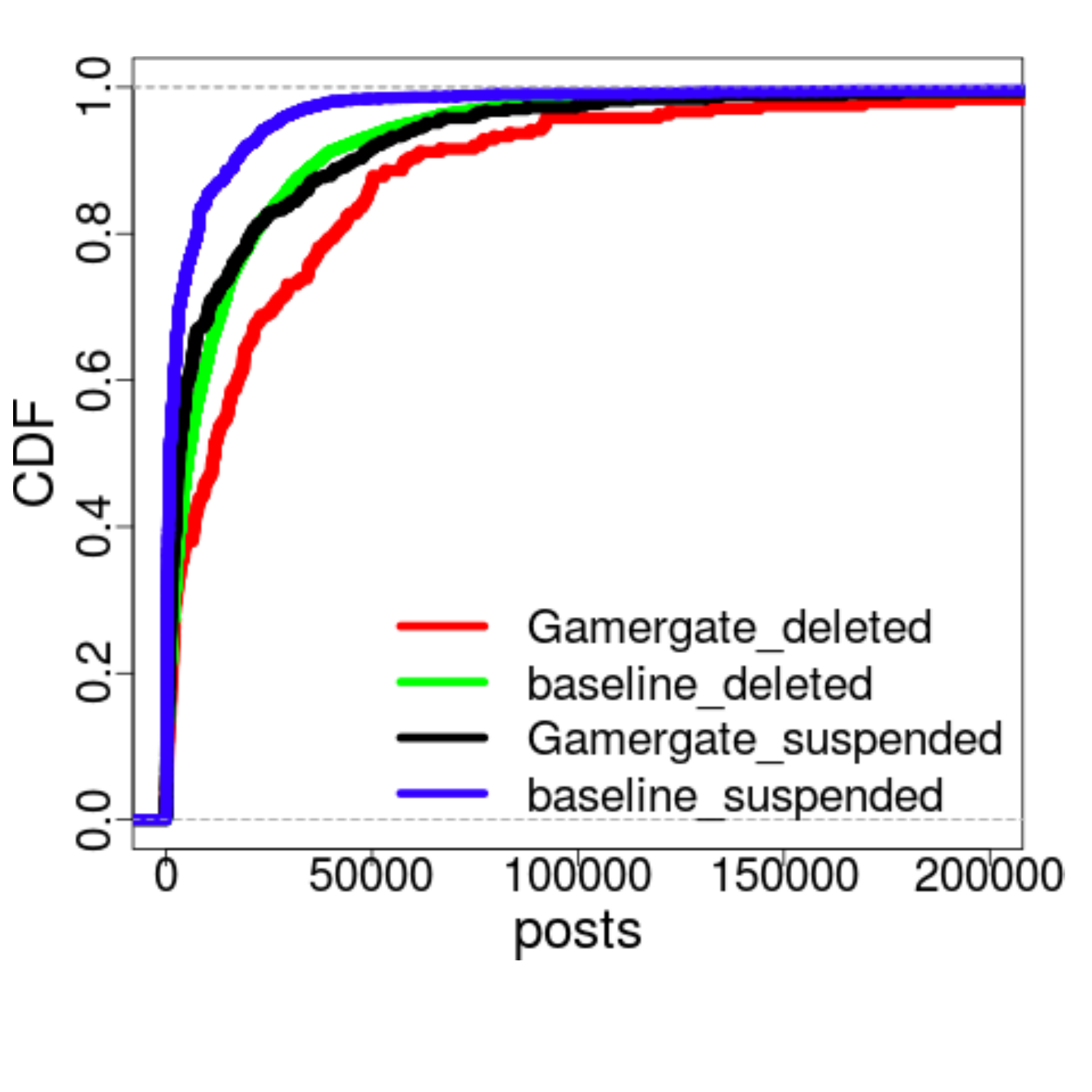}
		\captionsetup{font=scriptsize}
		\vspace{-0.95cm}
		\caption{Posts distribution.}
		\label{fig:sus_del_posts}
	\end{subfigure}	
	\begin{subfigure}[b]{0.23\textwidth}
		\includegraphics[width=\textwidth]{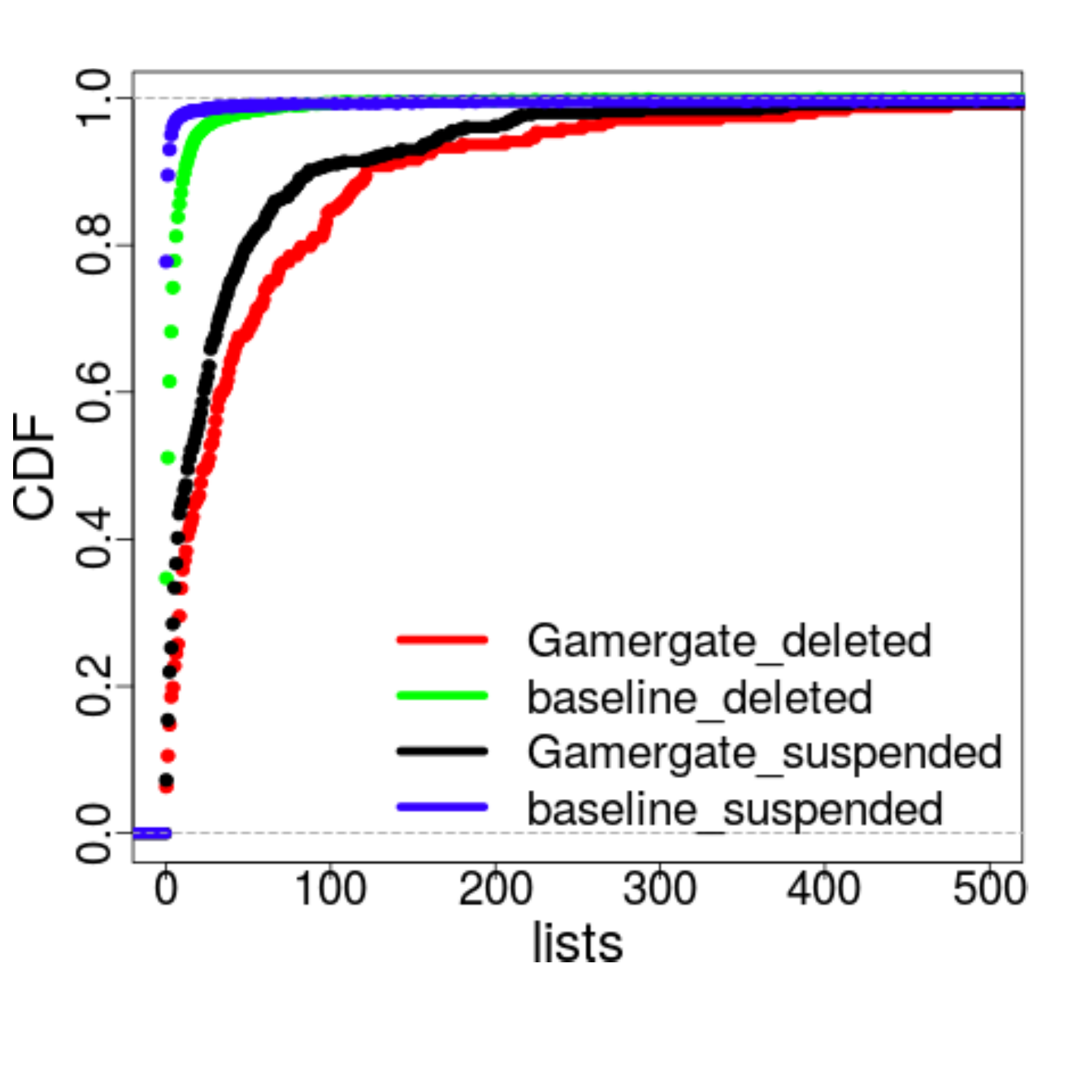}
		\captionsetup{font=scriptsize}
		\vspace{-0.95cm}
		\caption{Lists distribution.}
		\label{fig:sus_del_lists}
	\end{subfigure}	
	\begin{subfigure}[b]{0.23\textwidth}
		\includegraphics[width=\textwidth]{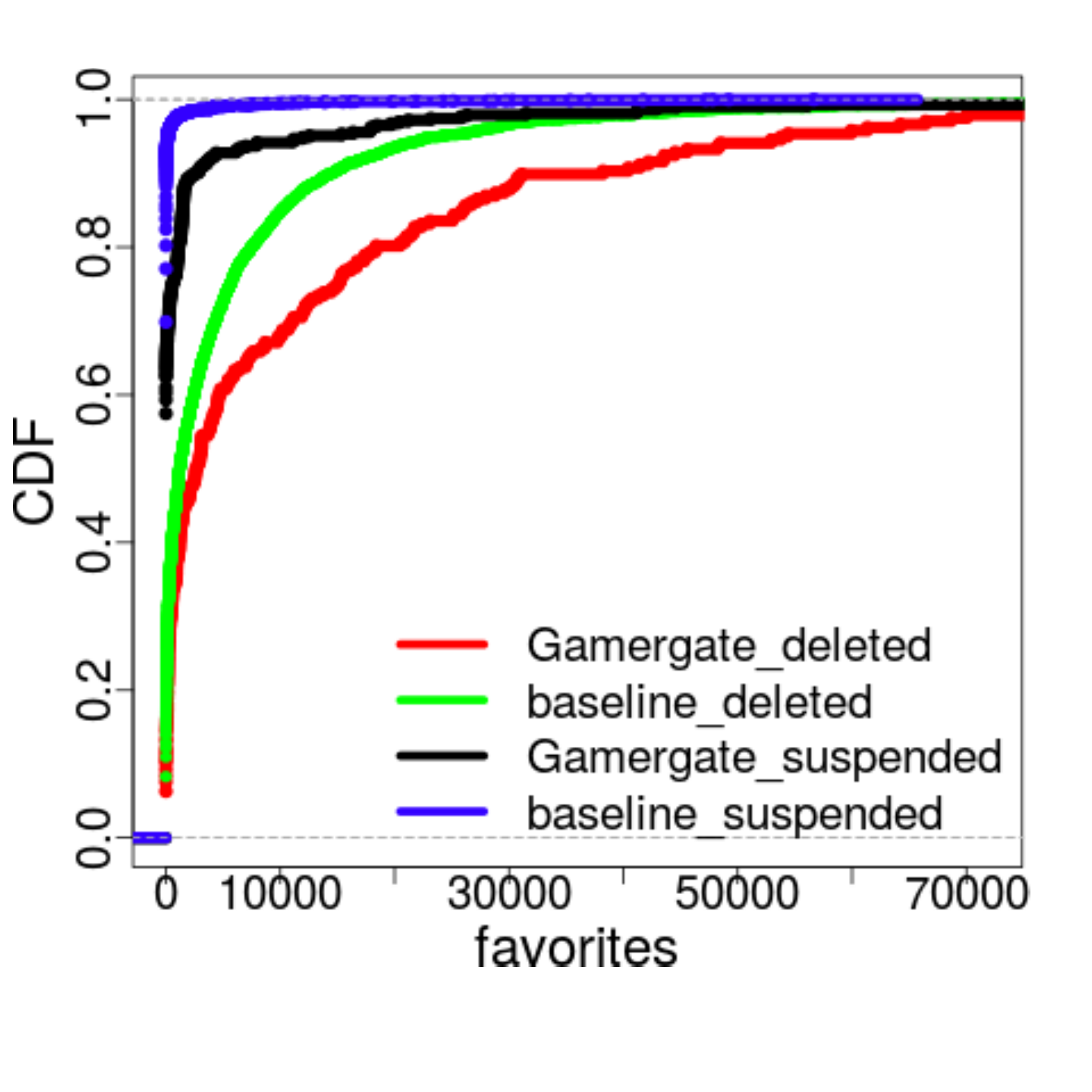}
		\captionsetup{font=scriptsize}
		\vspace{-0.95cm}
		\caption{Favorites distribution.}
		\label{fig:sus_del_favorites}
	\end{subfigure}	
	\vspace{-0.15cm}
	\caption{CDF plots for the suspended and deleted users considering activity attributes: (a) Account age, (b) Followers, (c) Friends, (d) Posts, (e) Lists, (f) Favorites.}
	\label{fig:sus_del_activity}
	\vspace{-0.3cm}
\end{figure}

Since users are suspended because their activity violates Twitter rules, and with the assumption that this detection system is consistent across users, we would expect GGers and baseline users to present similar behavior, or in some cases, we would expect GGers to be more extreme than baseline users.
On the other hand, users who delete their accounts could present a variety of behavioral attributes, as this decision is \emph{user-based}; i.e., there is a large number of confounding factors as to why the user decided to delete his account.
Based on Figures~\ref{fig:sus_del_emotional} and~\ref{fig:sus_del_activity}, we observe that there are substantial differences among the suspended baseline users and GGers, and the deleted baseline and GGers.

\descr{Sentiment, Emotions, and Offensive language.}
Concerning the emotional and sentiment attributes, we observe different behaviors.
For instance Figures~\ref{fig:sus_del_anger},~\ref{fig:sus_del_disgust}, and~\ref{fig:sus_del_offensive} show that suspended GGers are expressing more aggressive ($D = 0.76$, $p < 0.01$) and repulsive ($D = 0.23$, $p < 0.01$) emotions, and offensive language (use of hate words), in comparison to suspended baseline users.
Interestingly, suspended GGers also post more joyful tweets (Figure~\ref{fig:sus_del_joy}, $D = 0.44$ and $p < 0.01$), and even though $30\%$ of them post more negative sentiment tweets than baseline users, the rest of the suspended GGers are more positive than baseline suspended users (Figure~\ref{fig:sus_del_sentiment}, $D = 0.29$ and $p < 0.01$).
The posting of more aggressive and joyful tweets from suspended GGers contradicts the behavior observed earlier when studying the overall dataset of GGers (i.e., regardless of account status) which implies that such a deviation from the norm could be a reason for suspension.
Since extreme aggression, negative, and offensive language is abusive behavior, we would expect higher suspension rates for the GGers than baseline users.

In a similar fashion, we look at the deleted GGers and observe that they exhibit higher anger in their posted tweets than the deleted baseline users (Figure~\ref{fig:sus_del_anger}, $D = 0.39$ and $p < 0.01$), but lower than the suspended GGers.
They exhibit less joy (Figure~\ref{fig:sus_del_joy}, $D = 0.14$ and $p < 0.01$), but more sadness (Figure~\ref{fig:sus_del_sadness}, $D = 0.15$, and $p < 0.01$) and fear (Figure~\ref{fig:sus_del_fear}, $D = 0.13$ and $p < 0.01$) than the deleted baseline users and suspended GGers.
On the other hand, they tweet with sentiment which is more negative than suspended GGers and deleted baseline users (Figure~\ref{fig:sus_del_sentiment}, $D = 0.58$ and $p < 0.01$).
Finally, they type less in all uppercase than deleted baseline users (Figure~\ref{fig:sus_del_uppercases}, $D = 0.56$ and $p < 0.01$), but more than suspended GGers.
Based on these observations and in accordance with the higher expression of fear, it seems that deleted GGers are more emotionally introverted users, and might be deleting their accounts to protect themselves from negative behaviors/attention.

\descr{Age, Followers, and Friends.}
As far as the activity patterns, Figure~\ref{fig:sus_del_activity} shows that suspended and deleted GGers are more active overall than baseline users.
In particular, we observe (Figure~\ref{fig:sus_del_age}) that users who delete their accounts (GGers or baseline), have been on the platform longer than suspended users ($D = 0.51$, $p < 0.01$).
Surprisingly, for the limited amount of time their account was active, suspended GGers managed to become more popular and thus have more followers (Figures~\ref{fig:sus_del_followers}) and friends (Figure~\ref{fig:sus_del_friends}) than the suspended baseline users ($D = 0.64$ and $0.60$, $p < 0.01$ for both comparisons) and deleted GGers and baseline users.
The fact that the deleted users (GGers or baseline) have fewer friends and followers than suspended GGers, implies they have less support from their social network.
On the contrary, high popularity for suspended GGers could have helped them attract and create additional activity on Twitter and could be a reason for delaying the suspension of such highly engaging, and even inflammatory, users.

\descr{Posts, Lists, and Favorites.}
Figures~\ref{fig:sus_del_posts},~\ref{fig:sus_del_lists}, and~\ref{fig:sus_del_favorites} show the distribution of the number of posts, lists, and favorites, respectively, made by suspended GGers and baseline users, as well as deleted users.
Overall, we observe suspended GGers to be more active than baseline users, with more posts, higher participation in lists, and more tweets favorited ($D = 0.24$, $D = 0.74$, $D = 0.27$, $p < 0.01$ for all comparisons).
However, deleted GGers exhibit the highest activity in comparison to deleted baseline users ($D = 0.18$, $D = 0.58$, $D = 0.17$, $p < 0.01$ for all comparisons) as well as compared to suspended GGers.

Overall, deleted GGers appear to have been very active prior to their account deletion, have exhibited signs of distress and fear, and have shown, through their high posting activity, their anger, reduced joy, and negative sentiment.
However, their social network (ego-network of friends and followers) was either unsupportive, or just too small to provide the emotional support needed to block verbal attacks and aggression by other users who were involved in the GG controversy, and this overall hostile environment may have led them to delete their accounts.
Suspended users, however, managed to become highly popular in the platform in a short period of time and probably engaged in bullying and aggressive behaviors intense enough to lead to their suspension.

\subsection{Who should be suspended?}
\label{subsec:should-suspend}

In the previous paragraphs, we analyzed the behavior of GGers and baseline users, and compared them with respect to the status of their accounts (active, deleted, and suspended).
Furthermore, we observed that an important portion of the GGers remains active despite exhibiting, in some cases, abnormal behavior.
Here, we organize users in groups to understand what homogeneity or commonalities users have, e.g., if they all tweet with many hate words, negative sentiment, or anger.
By studying the heterogeneity of the identified groups, we then mark any diversity that users exhibit, and examine whether such a diversity could justify Twitter's tolerance against their abnormal behavior.

To group users who are highly similar over the available features studied, we use an unsupervised clustering method.
After the clustering task, we label the top 3 groupings created that cover the majority of users under a specific status.
We also investigate if the remaining clusters could be used to classify more users under the suspended status.

\descr{Clustering approach.}
Initially we extract both emotional and activity related attributes for the $33k$ users and proceed with a clustering process (separately on baseline and GG users, since we have seen a totally different behavior) in order to understand the commonalities behind Twitter's different statuses.
We use $K$-means~\cite{macqueen1967some}, an unsupervised learning algorithm, where each user in the dataset is associated to the nearest cluster centroid out of the $K$ clusters in total.
Each user $x$ is assigned to a cluster considering its distance from the $K$ cluster centroids $C$ as follows: $\argmin_{c_i \in C} dist(c_i, x)^2$.
In our case, $dist$ is the standard squared Euclidean distance in the $N$-dimensions used.
When all users are assigned to a cluster, the algorithm proceeds with a re-calculation of the $K$ new cluster centroids and a new binding of users to the nearest new centroids is made.
The re-calculation of the clusters' centroids is done by taking the mean value of the feature vector for users included in that centroid's cluster.
This process is completed when no change in cluster membership is observed, or a maximum number of iterations is reached.

\descr{Detecting the optimal number of clusters.}
In $K$-means the number of clusters to be extracted should be known a priori.
To find an appropriate number of clusters, one can run the $K$-means clustering algorithm for a range of $K$ values and compare the results with respect to compactness of clusters and distance between centroids.
A more sophisticated approach is to build upon the Expectation-maximization algorithm (EM)~\cite{gupta2011theory} which identifies naturally occurring clusters.
The EM algorithm is an efficient method to estimate the maximum likelihood in the presence of missing or hidden data.
Thus, given some observed data $y$, the EM algorithm attempts to find the parameters $\theta$ that maximize the probability:$ \theta = \argmax_\theta logp(y|\theta)$.
Then, for the unobserved or missing data $x$, we estimate $\theta$ that maximizes the likelihood, $l$, of $x$: $l(\theta) = \sum_{x}p(x,y;\theta)$.

\descr{Clustering tendency of Gamergaters.}
Considering the GG users based on the EM algorithm, we ended up with $3$ clusters for the emotional attributes and $8$ clusters for the activity-related attributes.
We see that some clusters are ``easily'' labeled due to the majority of users being one type of status.
Table~\ref{tbl:gg_emotions_clusters} (Table~\ref{tbl:gg_activity_clusters}) shows the distribution of the GGers in the $3$ ($8$) clusters which have been characterized as either active, deleted, or suspended using the Twitter status, and considering the emotional (activity)-based attributes.
As the GG dataset tends to contain a larger proportion of bullying and aggressive behavior phenomena, one would expect that based on the emotional-related features, the clustering results would be in better accordance to the Twitter status labels.
However, we observe that using the activity-related features results in those clusters better matching the Twitter applied status labels.

\begin{table}[!t]
\centering
\scalebox{0.8}{
\begin{tabular}{@{}l|c|rrr@{}}
\toprule
Status ->		&	Cluster	& \# active & \# deleted & \# suspended \\ \midrule
active		&	1		&	2,429      & 139        & 135      		    \\
deleted		& 	2		&	258        & 11         & 33           \\
suspended	&	3		&	1,615      & 87         & 260          \\ \bottomrule
\end{tabular}
}
\vspace{0.2cm}
\caption{Distribution of GG users in 3 clusters and the assigned label based on majority participation (emotional-related features).}
\label{tbl:gg_emotions_clusters}
\vspace{-0.2cm}
\end{table}

\begin{table}[!t]
\centering
\scalebox{0.8}{
\begin{tabular}{@{}l|c|rrr@{}}
\toprule
Status ->		&	Cluster	& \# active       & \# deleted       & \# suspended \\ \midrule
active		&	1	& 825             & 11               & 5                  \\
deleted		&	2	& 66              & 125              & 8                  \\
suspended	&	3	& 440             & 18               & 324                \\
			&	4	& 57              & 0                & 1                  \\
			&	5	& 757             & 27               & 56                 \\
			&	6	& 692             & 11               & 5                  \\
			&	7	& 725             & 32               & 22                 \\
			&	8	& 740             & 13               & 7                  \\ \bottomrule
\end{tabular}
}
\vspace{0.2cm}
\caption{Distribution of GG users in 8 clusters and the assigned label based on majority participation (activity-related features).}
\label{tbl:gg_activity_clusters}
\vspace{-0.2cm}
\end{table}

\descr{Clustering tendency for baseline users.}
We now perform the same analysis on the baseline users, looking for any differences of the suspension mechanism from GGers.
Here, the EM algorithm converged on $8$ clusters for \emph{both} the emotional and activity-related attributes.
Tables~\ref{tbl:baseline_emotions_clusters} and~\ref{tbl:baseline_activity_clusters} show these distributions, respectively.
We observe that for both feature sets, the cluster assigned the suspended label is clearly distinct, with substantially more users as members.
Deleted users are harder to fit: they do not seem to be primarily present in a single cluster, however, the activity-based features do seem to better cluster them.
This indicates that further analysis on deleted users should be conducted.

\begin{table}[!t]
\centering
\scalebox{0.8}{
\begin{tabular}{@{}l|c|rrr@{}}
\toprule
Status->		&	Cluster	& \# active & \# deleted & \# suspended \\ \midrule
active		&	1	&	4,999      & 1,501      & 658          \\
deleted		&	2	&	1,984      & 392        & 439          \\
suspended	&	3	&	4,200      & 690        & 3,832         \\
			&	4	&	3,333      & 373        & 134          \\
			&	5	&	1,308      & 358        & 120          \\
			&	6	&	1,030      & 169        & 162          \\
			&	7	&	1,525      & 133        & 257          \\
			&	8	&	433        & 85         & 71           \\ \bottomrule
\end{tabular}
}
\vspace{0.2cm}
\caption{Distribution of baseline users in 8 clusters and the assigned label based on majority participation (emotional-related features).}
\label{tbl:baseline_emotions_clusters}
\vspace{-0.2cm}
\end{table}

\begin{table}[!t]
\centering
\scalebox{0.8}{
\begin{tabular}{@{}l|c|rrr@{}}
\toprule
Status->		&	Cluster	& \# active & \# deleted & \# suspended \\ \midrule
active		&	1	&	6,885      & 1,121      & 651          \\
deleted		&	2	&	882        & 1,124      & 63           \\
suspended	&	3	&	4,942      & 574        & 3,765        \\
			&	4	&	1,580      & 156        & 74           \\
			&	5	&	2,733      & 594        & 78           \\
			&	6	&	858        & 51         & 51           \\
			&	7	&	142        & 24         & 2            \\
			&	8	&	787        & 57         & 989           \\ \bottomrule
\end{tabular}
}
\vspace{0.2cm}
\caption{Distribution of baseline users in 8 clusters and the assigned label based on majority participation (activity-related features).}
\label{tbl:baseline_activity_clusters}
\vspace{-0.4cm}
\end{table}

In general, and as expected, the clustering is not perfect in either of the two datasets: the clusters are fairly diverse with respect to users from the three statuses.
This is mainly because of two reasons:
(i)~the majority of users are active and since they exhibit a wide range of behaviors, they would be included in various clusters,
(ii)~some of these active users should probably be suspended, but the suspension mechanism failed to detect them.
Such users should be placed under evaluation for possible suspension.
To this end, we study the properties of the users included in each of the unnamed clusters (in emotional or activity-based clusterings), for baseline and GG users, and propose clusters which could be considered candidate for suspending users.

\begin{figure}[!t]
	\centering
	\begin{subfigure}[b]{0.23\textwidth}
		\includegraphics[width=\textwidth]{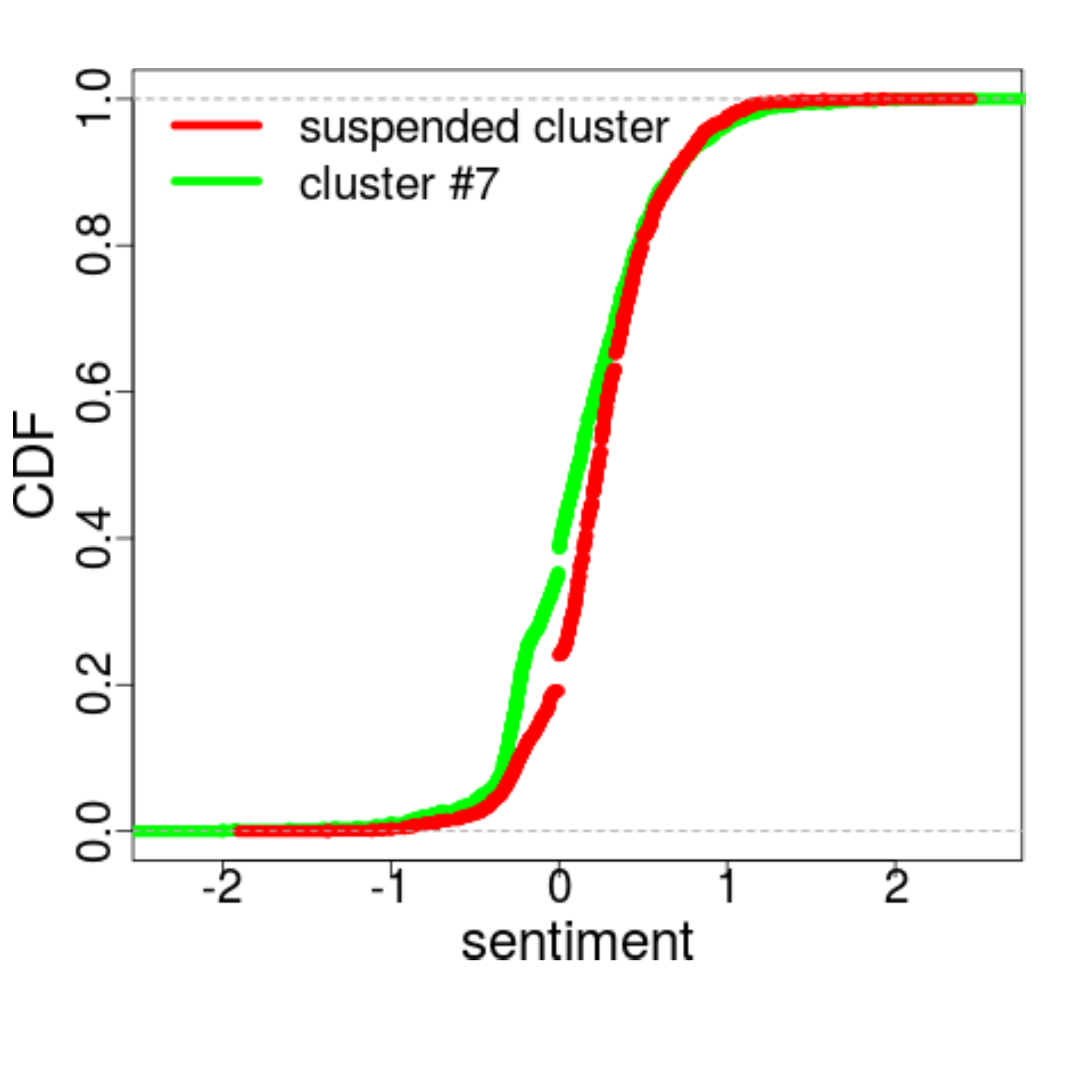}
		\captionsetup{font=scriptsize}
		\vspace{-0.95cm}
		\caption{Sentiment distribution.}
		\label{fig:baseline_unnamed_sentiment}
	\end{subfigure}
	\begin{subfigure}[b]{0.23\textwidth}
		\includegraphics[width=\textwidth]{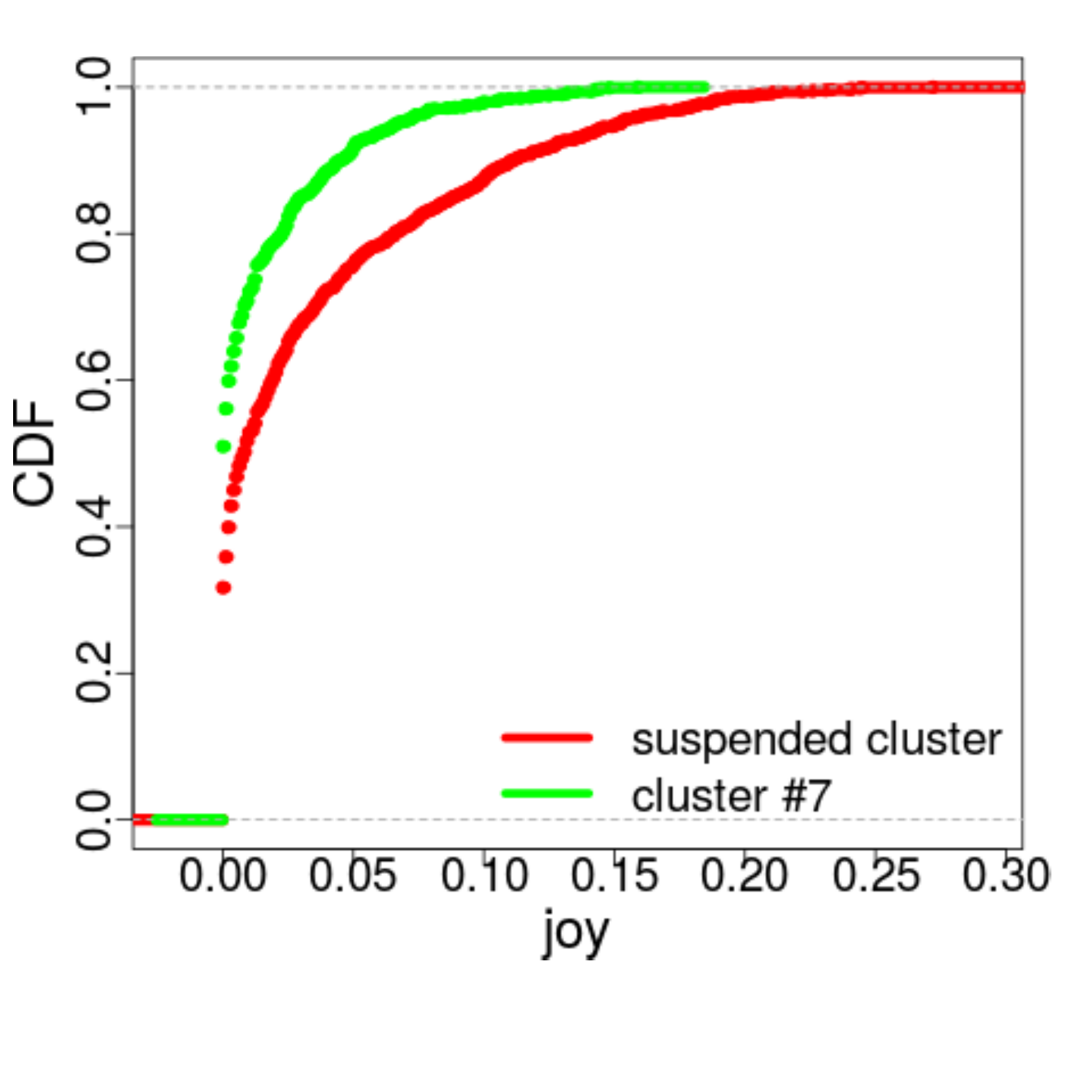}
		\captionsetup{font=scriptsize}
		\vspace{-0.95cm}
		\caption{Joy distribution.}
		\label{fig:baseline_unnamed_joy}
	\end{subfigure}
	\vspace{-0.15cm}
	\caption{CDF plots of baseline users for the suspended and an unnamed cluster considering the emotional attributes: (a) Sentiment, (b) Joy.}
\end{figure}

\begin{figure}[!t]
	\centering
	\begin{subfigure}[b]{0.23\textwidth}
		\includegraphics[width=\textwidth]{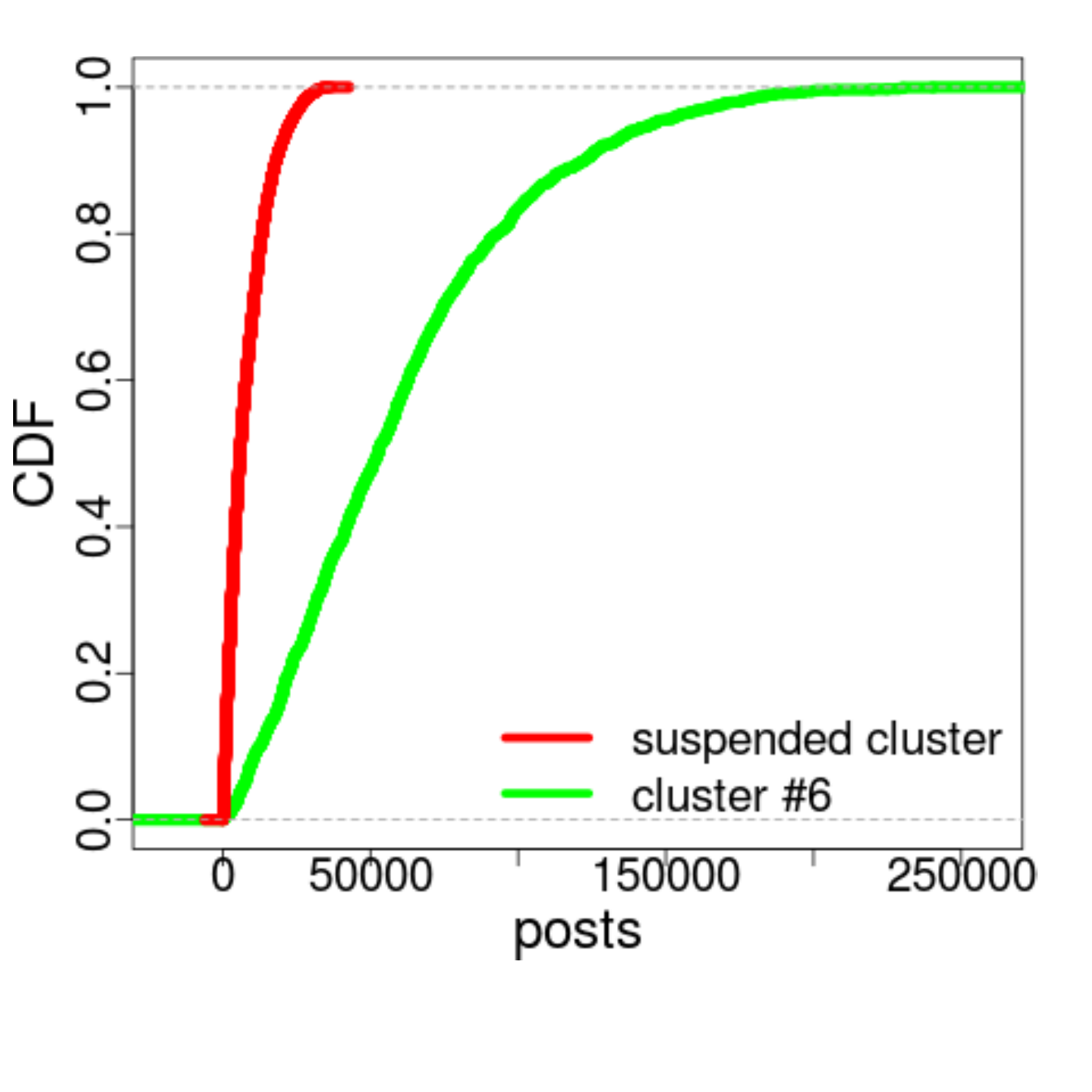}
		\captionsetup{font=scriptsize}
		\vspace{-0.95cm}
		\caption{Posts distribution.}
		\label{fig:baseline_unnamed_posts}
	\end{subfigure}
	\begin{subfigure}[b]{0.23\textwidth}
		\includegraphics[width=\textwidth]{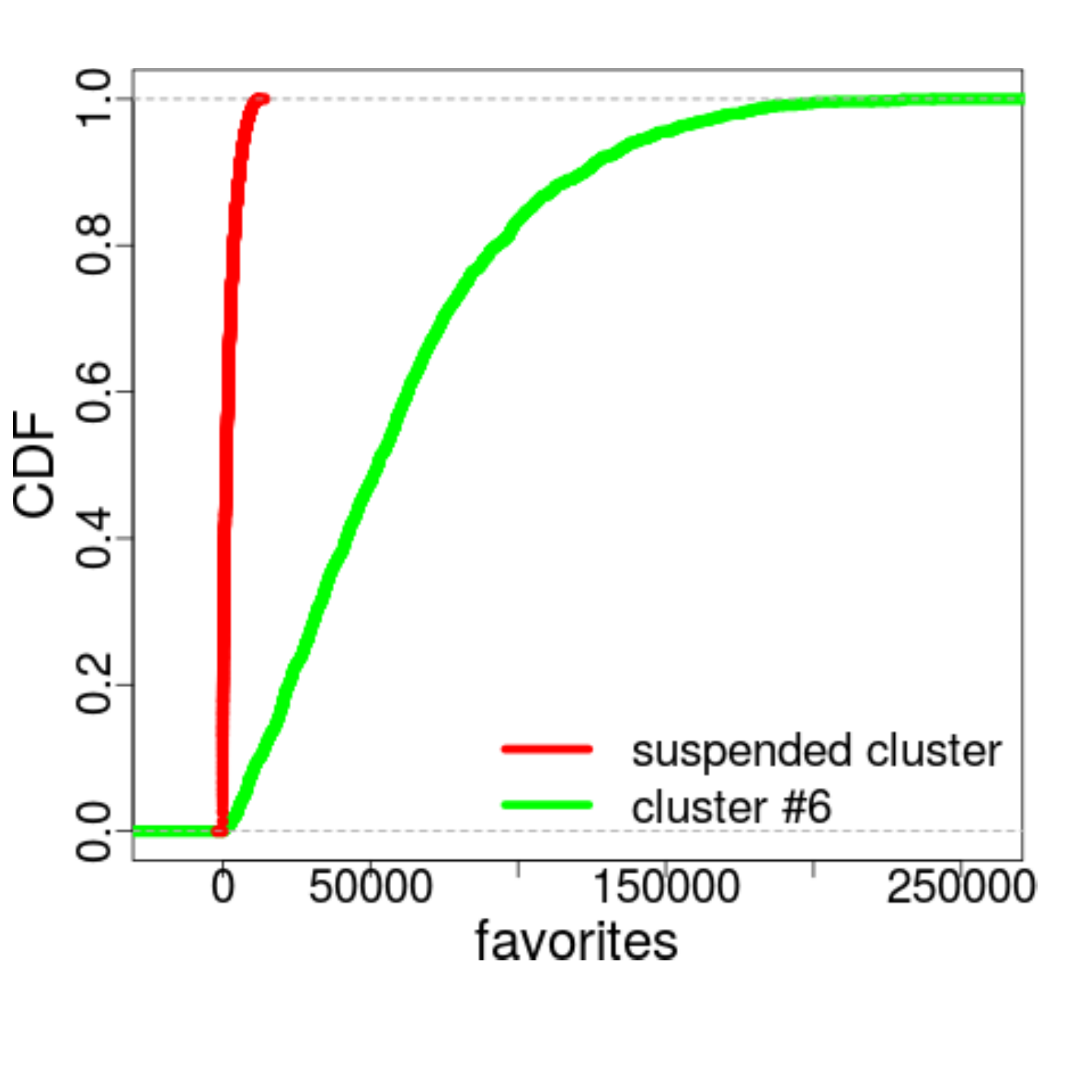}
		\captionsetup{font=scriptsize}
		\vspace{-0.95cm}
		\caption{Favorites distribution.}
		\label{fig:baseline_unnamed_favorites}
	\end{subfigure}
	\begin{subfigure}[b]{0.23\textwidth}
		\includegraphics[width=\textwidth]{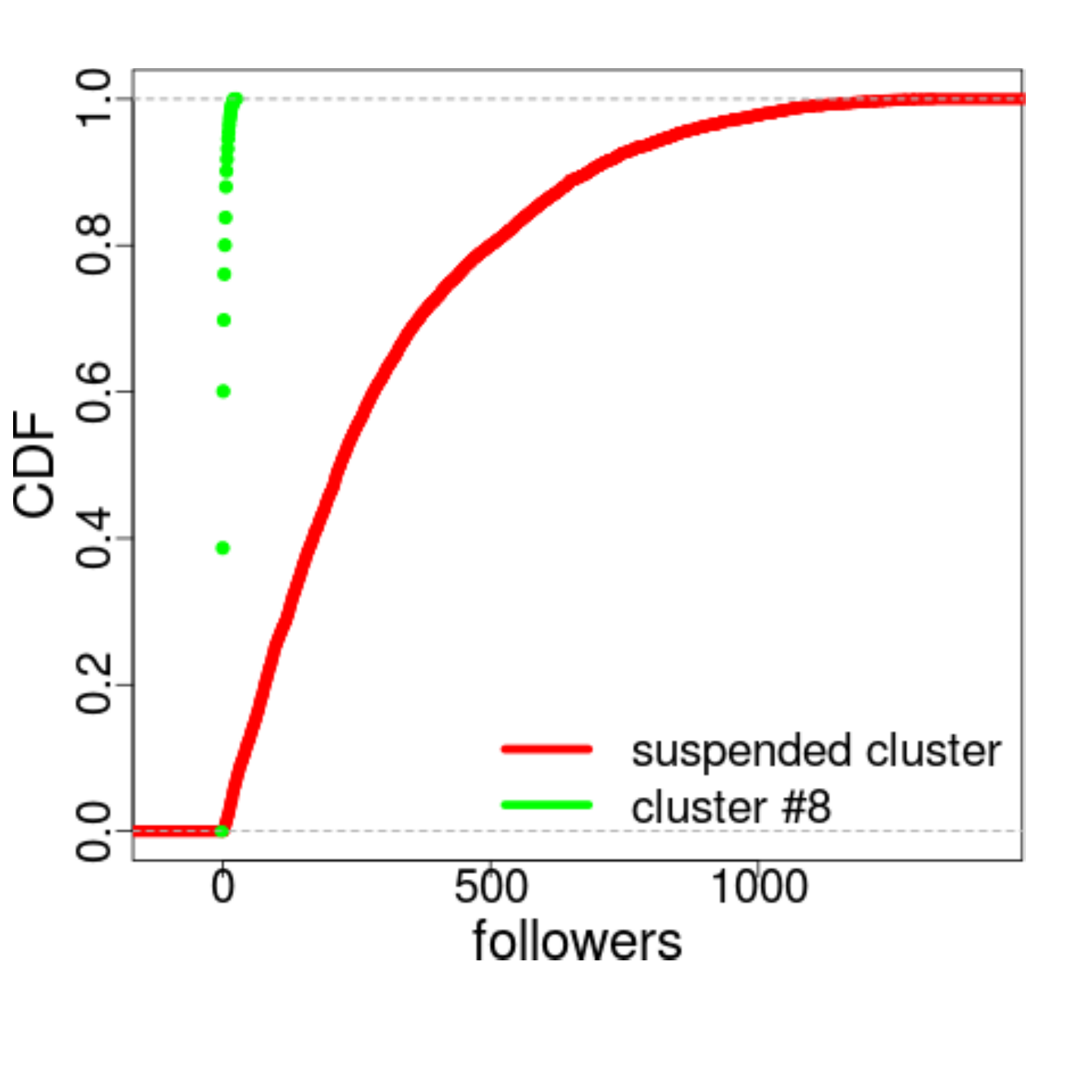}
		\captionsetup{font=scriptsize}
		\vspace{-0.95cm}
		\caption{Followers distribution.}
		\label{fig:baseline_unnamed_followers}
	\end{subfigure}
	\begin{subfigure}[b]{0.23\textwidth}
		\includegraphics[width=\textwidth]{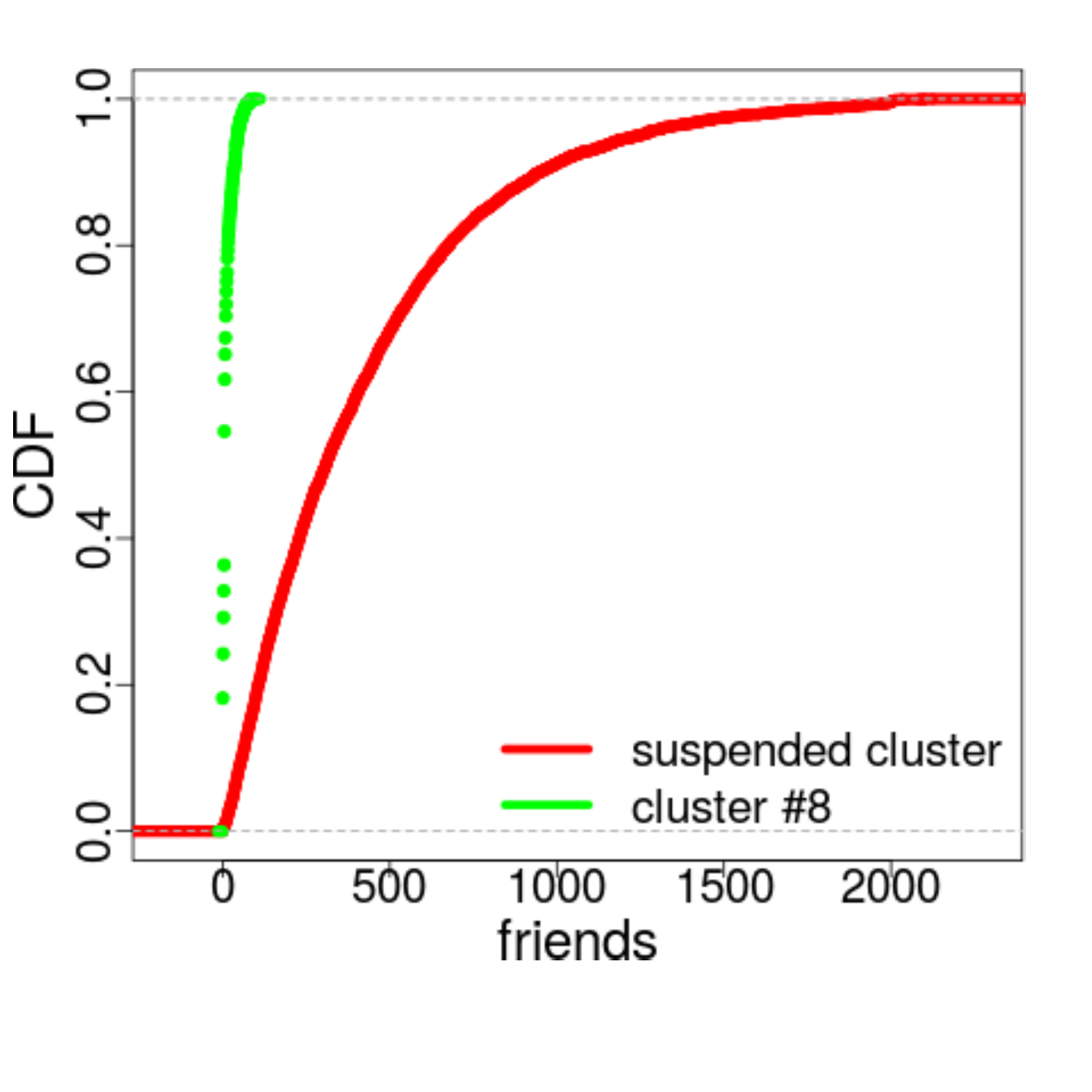}
		\captionsetup{font=scriptsize}
		\vspace{-0.95cm}
		\caption{Friends distribution.}
		\label{fig:baseline_unnamed_friends}
	\end{subfigure}
	\vspace{-0.15cm}
	\caption{CDF plots of baseline users for the suspended and an unnamed cluster considering the activity attributes: (a) Posts, (b) Favorites, (c) Followers, (d) Friends.}
	\vspace{-0.3cm}
\end{figure}

For instance, studying the unlabeled clusters of Table~\ref{tbl:gg_activity_clusters}, there is the cluster \#5 (with $757$ active, $27$ deleted, and $56$ suspended users) where GGers show similar activity patterns to those of the suspended cluster indicating that there are active users who could be possible candidates for suspension.
These users are similar to those of the suspended cluster: their accounts are pretty old and exhibit intense activity in terms of tweet posting ($mean = 23,664$, $median = 17,510$).
Also, they show similar patterns in terms of the favorited tweets and lists with the GGers of the suspended cluster.
Quite suspicious is the unlabeled cluster \#7 with $1,525$ active, $133$ deleted, and $257$ suspended users in Table~\ref{tbl:baseline_emotions_clusters}: the cluster members show signs of negative behavior by using offensive language and negativity in their tweets (Figure~\ref{fig:baseline_unnamed_sentiment}) and lower levels of joy (Figure~\ref{fig:baseline_unnamed_joy}).

Another unnamed cluster (\#6) which could be flagged as suspicious for suspension is the one with $858$ active, $51$ deleted, and suspended users in Table~\ref{tbl:baseline_activity_clusters}.
Here, both the suspended and \#6 clusters show similar activity in terms of list participation.
Quite interestingly, even though users in these clusters tend to have similar account age on Twitter, there are important differences in the number of their posted (Figure~\ref{fig:baseline_unnamed_posts}) and favorited (Figure~\ref{fig:baseline_unnamed_favorites}) tweets.
Such a disproportionality in the number of posted/favorited tweets (i.e., quite increased activity) and their lifetime on Twitter could be an indication of spam users.
Finally, focusing again on the same set of clusters (Table~\ref{tbl:baseline_activity_clusters}), cluster \#8 shows abnormal and consequently suspicious behavior.
The majority of its users have been suspended, but it also includes a lot of active users.
If we compare the popularity of users in cluster \#8 with the users from the suspended cluster, we find the suspended cluster users being more popular in terms of their followers (Figure~\ref{fig:baseline_unnamed_followers}) and friends (Figure~\ref{fig:baseline_unnamed_friends}).
However, the users in cluster \#8 have posted a relatively large number of tweets (the mean and standard deviation values are $547.15$ and $640.63$, respectively), considering their short lifetime. 
Such ``strange'' behavior could be indicative of spammer accounts.

\subsection{Emulating the suspension engine}
\label{subsec:status-classification}

Having gained an overview of the homogeneity or commonalities users have in accordance to their Twitter status, here we investigate if the features we have analyzed so far are meaningful and correlated with account statuses, and more importantly, if they can be used to automatically classify users.
To this end, we perform a supervised classification task using the three statuses as labels in an attempt to emulate the Twitter suspension engine.
We study the two types of users (GGers and baseline users) separately to understand if such features are more predictive of one or the other.

For the classification task, we test several tree-based algorithms as we find them to perform best (J48, LADTree, LMT, NBTree, Random Forest (RF), and Functional Tree).
Overall, tree-based classifiers are built from three types of nodes:
(i)~the \textit{root} node, with no incoming edges, 
(ii)~the \textit{internal} nodes, with just one incoming edge and two or more outgoing edges, and
(iii)~the \textit{leaf} node, with one incoming edge and no outgoing edges.
The root and internal nodes correspond to feature test conditions that separate data based on their characteristics, while the leaf nodes correspond to the available classes.
In the end, we select RF~\cite{rokach2014data}, as it achieved the best results with respect to time for training without overfitting the dataset.
We test the two categories of features (emotional and activity-based) separately, as well as combined.
Based on Tables~\ref{tbl:gg_results_classification} and~\ref{tbl:baseline_results_classification} in both the GG and baseline datasets, we observe that by considering the activity-related features, the precision, recall, and ROC (weighted area under the ROC curve) values are always higher at both the class level and overall across classes.
Adding all features together the scores are a little better than the emotional-related features (omitted due to space).

We remark that this classification task is not ideal for two main reasons:
(i)~we only use a subset of data and extract a subset of features from the ones that Twitter has available for making decisions with its status mechanism, and
(ii)~we only use a fairly simple, but robust, classification algorithm to attempt this task.
We suspect that Twitter computes many more features per user to assess their behavior, as well as using highly sophisticated algorithms for user suspension.
However, given these caveats, we show that it is possible to approximate the status mechanism, and perform very well with respect to standard machine learning metrics: we achieve $0.7-0.91$ precision, $0.76-0.92$ recall, and $0.75-0.89$ ROC.
From these preliminary results, we conclude that our features are meaningful in studying such user behaviors, and probably useful in detecting what status a user should be given by Twitter.

\begin{table}[!t]
    \vspace{1.7mm}
    \begin{subtable}[t]{.5\linewidth}
      \captionsetup{font=scriptsize}
      \centering
      	\scalebox{0.77}{
        \begin{tabular}{l|lll}
        \hline
				& \textbf{Prec.}	& \textbf{Rec.} & \textbf{ROC} \\ \hline
        active		& 0.898		& 0.982         & 0.747        \\ 
        deleted		& 0.667		& 0.008         & 0.550        \\ 
        suspended	& 0.669             & 0.407         & 0.865        \\ \hline
        overall (avg.)	& 0.867             & 0.886         & 0.747  	   \\ \hline
        \end{tabular}
        }
        \vspace{0.1cm}
              \caption{Emotional-related features}
      \label{tbl:gg_emotional_classification}
    \end{subtable}%
    \begin{subtable}[t]{.5\linewidth}
      \centering
       \captionsetup{font=scriptsize}
        \scalebox{0.77}{
        \begin{tabular}{l|lll}
        \hline
                       & \textbf{Prec.} & \textbf{Rec.} & \textbf{ROC} \\ \hline
        active         & 0.937          & 0.973         & 0.886        \\ 
        deleted        & 0.725          & 0.489         & 0.804        \\ 
        suspended      & 0.742          & 0.591         & 0.925        \\ \hline
        overall (avg.) & 0.910          & 0.917         & 0.886        \\ \hline
        \end{tabular}
        }
        \vspace{0.1cm}
        \caption{Activity-related features}
        \label{tbl:gg_activity_classification}        
    \end{subtable} 
    \vspace{-0.2cm}
    \caption{Classification results based on the GG dataset.}
    \vspace{-0.3cm}
    \label{tbl:gg_results_classification}

\end{table}

\begin{table}[!t]
    \begin{subtable}[t]{.5\linewidth}
      \captionsetup{font=scriptsize}
      \centering
      	\scalebox{0.72}{
        \begin{tabular}{l|lll}
        \hline
                       & \textbf{Prec.} & \textbf{Rec.} & \textbf{ROC} \\ \hline
        active         & 0.756          & 0.946         & 0.742        \\ 
        deleted        & 0.197          & 0.022         & 0.674        \\
        suspended      & 0.803          & 0.598         & 0.882        \\ \hline 
        overall (avg.) & 0.692          & 0.755         & 0.761  	   \\ \hline
        \end{tabular}
        }
        \vspace{0.1cm}
      \caption{Emotional-related features}
      \label{tbl:baseline_emotional_classification}
    \end{subtable}%
    \begin{subtable}[t]{.5\linewidth}
      \centering
       \captionsetup{font=scriptsize}
        \scalebox{0.72}{
        \begin{tabular}{l|lll}
        \hline
                       & \textbf{Prec.} & \textbf{Rec.} & \textbf{ROC} \\ \hline
        active         & 0.806          & 0.943         & 0.826        \\ 
        deleted        & 0.570          & 0.248         & 0.806        \\ 
        suspended      & 0.892          & 0.718         & 0.937        \\ \hline
        overall (avg.) & 0.792          & 0.807         & 0.846        \\ \hline
        \end{tabular}
        }
        \vspace{0.1cm}
        \caption{Activity-related features}
        \label{tbl:baseline_activity_classification}
    \end{subtable} 
    \vspace{-0.2cm}
    \caption{Classification results based on baseline dataset.}
    \vspace{-0.3cm}
    \label{tbl:baseline_results_classification}

\end{table}

\section{Discussion and Conclusion}\label{sec:discussion}

In this paper we have performed a large-scale comparative study of abusive accounts on Twitter, aiming to understand their characteristics and how they differ from regular accounts.
Specifically, we focused on a Twitter dataset revolving around the Gamergate controversy which led to many incidents of cyberbullying and cyberaggression on various gaming and social media platforms.
We studied the properties of users tweeting about GG, the content they post, and the differences in their behavior compared to typical Twitter users.
We found that users involved in this controversy were existing Twitter users that were probably drawn to the controversy.
In fact, their familiarity with Twitter could be the reason that GG exploded in the first place.
We also discovered that while the subject of their tweets is seemingly aggressive and hateful, GGers do not exhibit common expressions of online anger, and in fact primarily differ from typical users in that their tweets are less joyful.
This aligns with the viewpoint of the GG supporters who claim that they never agreed to the aggressive methods used in this campaign~\cite{Mortensen2016}, which can result in a confusing expression of anger manifestation.
GGers tend to be organized in groups, and in fact they participate also in face-to-face meetings to create stronger bonds, which also reflects on the higher number of followers and friends they have in relation to typical users, despite their seemingly anti-social behavior.
Also, we discover that GGers are seemingly more engaged than typical Twitter users, which is an indication as to how and why this controversy is still ongoing.

To better understand how these abusive users are handled by Twitter, we performed an in-depth analysis of the status of accounts posting about GG and typical Twitter users.
Surprisingly, we found that GGers are disproportionally \emph{not suspended} with respect to random users, which is rather unexpected given their hateful and aggressive postings.
Therefore, we investigated  users' properties with respect to their account status to understand what may have led to suspension of some of them, but not all of them.
Even though suspended GGers are expressing more aggressive and repulsive emotions, and offensive language than random users, they tend to become more popular and more active in terms of their posted tweets.
This popularity could be the reason for the delayed suspension from the Twitter mechanism, a situation that seems to have changed lately, considering the new actions taken by Twitter itself, e.g.,~\cite{CNNtech,independent}.

We also studied the GG users who deleted their account.
These users demonstrate the highest activity in comparison to other users (deleted or suspended).
Overall, such deleted users exhibit signs of distress, fear, and sadness, and have probably showed these emotions through their high posting activity filled with anger, reduced joy, and negative sentiment.
We also found that such users have small social ego-networks, which may have been unsupportive or too small to help them deal with aggressive attacks by other GGers before deleting their account.

Finally, we performed an unsupervised machine learning analysis to detect clusters of users who, though currently active, could be considered for suspension as they exhibit similar behaviors with already suspended users.
Our findings are a first step towards understanding better, and at large-scale, the behavior of abusive users in online social media such as Twitter, their victims and what may have led them to delete their account, and propose supervised methods to detect suspicious users whose accounts should be evaluated for suspension.
As part of future work, we plan to perform a more in-depth study of the Gamergate controversy and further compare it with other organized groups that exhibit online aggressive and abusive behaviors.

\section{Acknowledgement}{
This research has been fully funded by the European Commission as part of the ENCASE project  (H2020-MSCA-RISE of the European Union under GA number 691025).}

\bibliographystyle{abbrv}
%\bibliography{bibfile}

\end{document}